\documentclass[article]{elsarticle}
\usepackage{graphicx}

\usepackage{amsmath}
\usepackage{algorithm,algorithmic}
\usepackage{hyperref}
\usepackage{subfig}
\usepackage{multirow}
\usepackage{amssymb}

\begin{document}

\begin{frontmatter}
\title{Detecting malicious logins as graph anomalies}

\author{Brian~A.~Powell}
\address{Johns Hopkins University Applied Physics Laboratory, Laurel, MD 20723}
\ead{brian.a.powell@jhuapl.edu}


\begin{abstract}
Authenticated lateral movement via compromised accounts is a common adversarial maneuver that is challenging to discover with signature- or rules-based intrusion detection systems.  In this work a behavior-based approach to detecting malicious logins to novel systems indicative of lateral movement is presented, in which a user's historical login activity is used to build a model of putative ``normal'' behavior.  This historical login activity is represented as a collection of daily login graphs, which encode authentications among accessed systems with vertices representing computer systems and directed edges logins between them. We devise a method of local graph anomaly detection capable of identifying unusual vertices that indicate potentially  malicious login events to the systems they represent. We test this capability on a group of highly-privileged accounts using real login data from an operational enterprise network. The method enjoys false positive rates significantly lower than those resulting from alerts based solely on login novelty, and is generally successful at detecting a wide variety of simulated adversarial login activity.   
\end{abstract}


\end{frontmatter}


\section{Introduction}
A healthy mindset in cybersecurity is that your enterprise will eventually be breached. Once inside the network, the adversary will get quiet---they will live off the land stealing credentials and moving between systems via authenticated access. They won't be so foolish as to launch noisy exploits or deploy malware that could set off alarms.  They will do their best to look like authorized users doing legitimate things, dwelling in the network for weeks or months as they quietly root out and steal your critical data.  

Most commercial solutions that provide network intrusion detection and prevention at the perimeter and on endpoint systems do not adequately protect against lateral movement; those offerings that do tend to take considerable time to configure and test, and are often proprietary black boxes (e.g. Windows ATA or FireEye TAP).  That's not to say that organizations don't have defensive strategies making use of internal sensor data and logs, but these tend to involve rule-based indicators.  The trouble with static rules is that they apply to known, explicit patterns of activity, that is, the threat indicators must have been previously observed.  Furthermore, there is a sort of selection bias at play in which only those threat activities that possess stable and recognizable signatures are ever even considered for detection.   Alas, authenticated lateral movement does not follow an explicit or recurrent pattern: the sequence and tempo of accesses, and the tools used and the systems targeted, vary according to the whims, tactics, and plans of the adversary.

Lateral movement via compromised accounts does have a common tell: the adversary's pattern of accesses need not include systems visited very often, if ever, by the compromised account.  Such {\it novel logins} might be a sign that a credential has been stolen or an authorized user is up to no good. But for privileged accounts with broad access like network administrators, novel logins might also be normal and common.  Since such accounts are precisely those most often targeted by attackers, this particular indicator will generate many false positives, potentially swamping a true signal of adversarial activity.  The difficulty here is that novelty does not imply malice.  

Given that the adversary is a rational agent intent on achieving some objective in an orderly, stealthy manner, we expect that there {\it are} patterns in lateral movement, but they are not known {\it a priori} by defenders. Instead, they must be inferred from observed activities---the problem is that data on generic adversarial lateral movement is hard to come by in sufficient quantities.  On the other hand, organizations {\it do} have lots of data tracking and characterizing {\it normal} activities on their networks.  Appropriate paradigms for cybersecurity analytics are therefore unsupervised and one-class learning, in which a sample of data describing normal activities is used to generalize and model it: anything that deviates significantly from what the model considers normal is deemed anomalous.  In this way, we can use unsupervised learning to go beyond simple alerts based on novelty. 

The picture of a given user's remote login activity over a period of time takes the form a {\it weighted, directed graph}: vertices are systems and directed edges are logins between them.  These {\it login graphs} summarize a user's authentication activity over some period of time, say, one day.  We will write the login graph of a given user, $u$, for day $i$ as $G^u_i$.  Over the course of several weeks, each user will have a sequence of login graphs, $\mathcal{G}^u = \{G^u_1,G^u_2,...,G^u_m\}$ for $m$ days of activity.  If this user's account gets compromised on one or more of these days, it is possible that the corresponding login graphs will appear anomalous relative to others, as the adversary performs authentications differing from this user's normal pattern of behavior.  

The scenario we have in mind is daily testing of a group of privileged user's login graphs for signs of compromise.  Each user has a history of presumed normal login graphs, $\mathcal{G}^u$, beginning $m$-days ago and ending on the day of testing,  $\mathcal{G}^u = \{G^u_{-m+1},G^u_{-m+2},...,G^u_0\}$.  Given the likelihood that adversarial lateral movement will include authentications to {\it novel systems}, we only wish to test $G^u_0$ if it contains logins to systems not present in the user's login record included in $\mathcal{G}^u - \{G^u_0\}$.  Specifically, novel systems correspond to vertices $v_{0,j} \in G^u_0$ with a label $j$ (IP address, host name, or some other unique identifier) not appearing in $\mathcal{G}^u - \{G^u_0\}$. 

There has been considerable research on the problem of discovering anomalies in time-evolving sequences of graphs \cite{Aggarwal14,Akoglu,Ranshous}.
These methods summarize graphs using global \cite{Pincombe,Gaston,Ide,Wilson,Berlingerio} or substructure \cite{Nobel,Macindoe,Sun07,Vishwanathan,Papadimitriou,Moriano} characteristics and either study how these characteristics change over time, or perform clustering or classification \cite{Li12,Gamachchi1,Kaiafas} on these characteristics in an attempt to identify anomalous graphs.  Alternatively, difference metrics have been developed that quantify the extent to which pairs of graph are unalike, and anomalous graphs can be identified as those that differ the most from all others \cite{Pincombe,Shoubridge99,Shoubridge02,McWherter,Peabody,Bunke04,Bunke06,Koutra}.  While these methods could be applied to determine if $G^u_0$ is anomalous, they seem particularly suited to users whose normal login graphs are relatively similar from day-to-day (such that a sense of ``normal'' can be applied to the entire login graph), and for adversarial activities that notably disrupt this behavior, resulting in login graphs that look globally different from normal graphs.  In our analysis, we don't necessarily expect (nor do we observe) users having similar graphs from day-to-day, and we also anticipate that the adversary might be so quiet as to cause only tiny modifications to a compromised user's daily login graph.  These considerations suggest that instead of looking to identify entire login graphs as anomalous, we instead seek a more local measure of graph anomaly, ideally at the vertex level so that individual systems can be assessed.

If graph vertices can be tracked over time, techniques from time series analysis can be applied to detect outliers \cite{Ide,Priebe,Akoglu3,Neil,Rossi,Wang,Palladino}.  These approaches rely on {\it node correspondence}, that is, there must be a vertex with a given label present at each time step so that its history can be traced out over time.  In our application, however, novel vertices have no history---the test day is the first time they appear in the user's login record.  We therefore cannot adopt a notion of anomalous vertex as one that appears different relative to its own history.  Instead, we consider whether novel vertices look like {\it other} vertices with respect to local and neighborhood graph properties, {\it i.e.} its pattern of connections to other vertices in its login graph. Let us write\footnote{For brevity we omit the superscript $u$ with the understanding that we are analyzing a single user's login graphs.} one such novel vertex as $v_{0,j^*} \in G_0$.  The vector-valued function ${\bf f}(v_{0,j^*})$ quantifies the local and neighborhood properties of the vertex (including, as we shall see, measures like degree, eigencentrality, and eccentricity), and we wish to compare ${\bf f}(v_{0,j^*})$ with all of the other vertices in $\mathcal{G}$, that is ${\bf f}(v_{i,k})$ for $-m+1 \leq i \leq 0$ and all $k \leq |G_i|$.  An important aspect of this approach is that it is not chronological and labels are not used: we are simply comparing the properties of a large collection of normal vertices, collected over all login graphs in $\mathcal{G}$, with those of the novel vertices we wish to test.

In this paper, we propose a new method of unsupervised detection of anomalous vertices in a sequence of graphs useful for discovering potentially malicious logins to novel systems on a computer network. Each vertex is summarized in terms of a set of graph centrality and topology measures, and then the set of all vertices are taken together and compressed to a lower-dimensional feature space. The purpose of this transformation is to assign each vertex to one of a small number of roles \cite{Rossi,Henderson2}; we can then measure how well each vertex fits into its role using the transformation's reconstruction error. By analyzing the statistics of the reconstruction errors in the sample, outlier vertices can be identified, corresponding to systems that are accessed in ways atypical of the user in question.

We test this approach on a group of around 75 privileged users on a large, operational enterprise network.  This group includes users with considerable access on a large number of endpoint systems and servers---Domain Administrators, Helpdesk operators, and Desktop and Server administrators---precisely the kinds of high-value accounts that should be monitored for compromise. The graph sequence $\mathcal{G}^u$ is constructed for each user from authentication log data over a period of four weeks.  To validate this approach against adversarial logins, we simulate a varied set of {\it adversarial login graphs}, which are devised to be representative of a threat's lateral movement tactics.  For most users, models are successful at detecting a wide range of adversarial login activities while exhibiting low false positive rates of around 2\%.  We compare this approach with a variety of alternatives from the literature and find that it performs best for our application.
\section{Related Work}
This work applies graph anomaly detection to the problem of discovering malicious activities inside computer networks.  It therefore contrasts with and draws from prior work in each of these areas.  We discuss the relevant literature from each area separately below.
\subsection{Lateral Movement and Malicious Logins}
The focus of our work is the detection of malicious logins occurring over the course of adversarial lateral movement, which is the sequence of internal accesses achieved by an attacker after gaining a foothold within the network.  These accesses are authentications, and so attackers look like authorized users.  Studies that seek to detect malicious insiders---authorized users abusing their access for nefarious reasons---consider similar data sources and indicators, and so we include some of this research in our literature review.  

One approach to detecting unauthorized user activity, including lateral movement, is to examine host-based indicators like the use of certain programs, system calls, or terminal commands.    In \cite{Li}, profiles of how individual users interact with their computers over time (in terms of things like processor usage and number of windows open) are built and tested for anomalies in the one-class setting.  These profiles are useful for identifying masquerades, in which one profiled user begins to act like another. An approach that models a user's activities as a bipartite user-system graph is presented in \cite{Gamachchi1,Gamachchi2}.  Data from web accesses, system logons, removable media actions, file movement, email, and psychometric data are represented graphically, and each user's graph is summarized in terms of a small collection of measures, including degree, edge count, density, diameter, and number of peers.  One-class learning is used to identify anomalous users according to these measures.  System calls formed the basis of lateral movement and masquerade detection in \cite{Jha,Hofmeyr,Lee,Warrender}; command-line sequences were analyzed for malicious activity by \cite{Maxion,Schonlau,Lane,Davison,Tomonaga}. 


In contrast to these approaches, we are interested in analyzing only a user's pattern of login events to identify anomalies.  It does not matter what type of login it is, which tool was used, or what activities were conducted on the host before or after access.  Further, the above methods make use low-level host-based data (like system call traces and terminal histories) that are potentially too voluminous to be centrally collected in a SIEM.  In contrast, login patterns can be obtained from login event data ({\it e.g.} Windows Event Logs) which are widely available and useful for a range of security tasks.

A work quite germane to ours is \cite{Kent}, who analyze login graphs to identify compromised user accounts.  The setting for this study was a one-month long red team assessment at Los Alamos National Laboratory (LANL): login graphs like the ones we consider here were created for all users over this time period, and supervised learning was applied to a set of global features derived from these graphs to train a classifier to discriminate normal from compromised user accounts.  The primary difficulty with this study, at least from an implementation standpoint, is the use of supervised learning: organizations generally do not have reliable adversarial data on the relevant kinds of activities in sufficient quantity to build an accurate classifier.  Furthermore, discrimination was done at the user level, in comparison to other users, on graphs spanning several weeks of activity.  In an operational setting, this is long lag time between initial compromise and detection.  The approach we develop here instead uses unsupervised learning on normal data only, applied to login graphs spanning a single day.  Rather than comparing across users, we compare across login activities of a single user to spot anomalies.

Other supervised approaches include the work of \cite{Kaiafas}, which applies an ensemble of classifiers to learn features derived from the LANL authentication data.  These features are derived from bipartite graphs of user-system interaction, and include event timing and user variance characteristics.  A study focused solely on lateral movement via Remote Desktop Protocol (RDP) \cite{Bai} improves on \cite{Kaiafas} by taking into account the biased nature of malicious samples (red team events originated from a very small set of hosts). This highlights the general difficulty with training on malicious data: it tends to involve a small number of entities (hosts, users, or access techniques) that can easily bias learning if they are used as features.  The work of \cite{Bian} augments the LANL authentication data with features derived from flow records: their random forest classifier is competitive with \cite{Kaiafas} but they report better generalization in the face of novel samples. 

A semi-supervised learning-based lateral movement detector was explored in \cite{Chen18}.  Communications graphs representing inter-host connections were constructed from a variety of data including authentications, network connections, and DNS queries among hosts.  This approach employed network embedding to aggregate graph features, then a denoising autoencoder was used to extract the most representative features.  Autoencoders were then trained on these representations to identify lateral movement. Though the classifier performed well on balanced LANL data, the authors acknowledge that in the deployed environment where normal events are expected to overwhelm malicious activities, this method can be expected to generate many false positives.  Indeed, \cite{Bian} confirms this for the unbalanced data set.  Autoencoders with deep architectures were also applied to the LANL data in \cite{Holt}.  Features were based on the frequencies of connections with the same/different source and destination users, and similar features for systems. An interesting aspect of this work is that the autoencoder is trained on the lateral movement---the red team events from the data set---under the assumption that it's easier to profile the needle than the haystack.  While perhaps true, it makes the method reliant on substantial amounts of sufficiently generic adversarial data to make a robust classifier.  This approach therefore faces the same implementation challenges as those based on supervised learning. 

Unsupervised and one-class learning paradigms have also been explored.  In \cite{Siadati}, login patterns were extracted from user authentication histories using market basket analysis applied to user-source system-destination system triplets.  The method is able to recognize when users with certain attributes access systems of a certain type or role that is not a learned pattern from the user's history.  The limitation of this approach is that the adversary may well only access systems with roles within the compromised user's history, {\it e.g.} a database administrator might only be able to access database systems.  Further, on large enterprise networks obtaining the ground truth on server roles is likely to be a manually intensive process.  

Prior works most similar to ours apply unsupervised or one-class learning to graphical models of user login activity.  In \cite{Bohara}, a network-wide communication graph is constructed including all traffic between all nodes.  The approach assumes that command and control and lateral movement monitors, technologies capable of scoring systems according to their probability of compromise and charting out causal connection paths through systems, like \cite{Fawaz}, are in place across the network.  Features include command and control score and four metrics characterizing the distribution of lateral movement chains, which are reduced via principal component analysis and clustered to find anomalies.  This approach finds only relatively long lateral movement chains (many infected hosts) with reasonable accuracy ($\approx 5\%$ false positive rate for $>4000$ infected systems) with degraded performance for smaller chains. In contrast, we seek to discover much smaller infections with as few as one malicious login. 
 
In \cite{Chen}, a bipartite graph of users and systems with which they interact is constructed from access logs.  Intruders are found by removing users one-by-one and comparing the user-system affinity matrices before and after removal.  A similarity is score is computed and thresholded to identify anomalous users.  This unsupervised approach has applications to collaborative information systems, for which groups of users are expected to assume common roles.  In this work, we consider a group of several dozen privileged accounts, like Domain Administrators and Helpdesk operators; however, despite generally similar roles, they do not exhibit similar login behaviors.  And, as with \cite{Kent}, we wish to discover anomalies within a single user's history rather than in comparison with other users.     

The work of \cite{Eberle09,Eberle10} looks for anomalies in login graphs indicative of malicious insider activity. Three types of anomalies are considered: graph modifications, insertions, and deletions.  The proposed algorithm, graph-based anomaly detection \cite{Eberle07,Davis}, looks for common substructures within a graph or sequence of graphs by compressing subgraphs and scoring them by their minimum description length (MDL). The least compressible structures are considered anomalous, following the proposal of \cite{Nobel}. This approach, while not at the vertex-level, is potentially quite relevant to our problem and we compare it to our approach later in the paper.  Community-level anomaly detection was applied to login graphs in \cite{Moriano}, in which communities were examined over time and a detection measure based on changes in inter-community edges was used to find anomalies.  In \cite{Ding}, vertices associated with multiple communities in graphs of network traffic among systems were identified as potential intrusions.

In \cite{Heard}, network-wide authentication graphs were constructed using the LANL data.  For each destination system, the distribution of source system connections is modeled as a Dirichlet process and the p-value of each connection is computed.  The p-values of each source system, across all edges, are then combined and a total measure of ``surprise'' for that source is computed.  Another work that applies probabilistic modeling is \cite{Turcotte},  which treats interaction counts between users and systems as a Poisson process.  Anomalies are those interaction pairs that have threshold p-values.  These methods cannot be applied to novel systems, however, since the corresponding vertices lack a historical distribution against which to assess anomalous behavior.

Lastly, we mention the work of \cite{Amrouche} who analyze the paths in authentication graphs that terminate at known malicious events.  This approach is indicated for root cause analysis, and so is investigative rather than detective.  
\subsection{Graph Anomaly Detection}
Graph anomaly detection is a diverse field with techniques developed for many applications, including fraud detection, social network analysis, intrusion detection, and fault tolerance in complex networks (see \cite{Aggarwal14,Akoglu,Ranshous} for reviews).  There are two broad approaches with potential relevance to our problem: detection of anomalies within single graphs, in which vertices and substructures are deemed anomalous with respect to other vertices and substructures within the same graph; and  the detection of anomalous graphs, considered as a whole, in comparison with a collection of other graphs.  A third approach, which seeks to identify vertices and substructures as temporal anomalies over time-ordered sequences of graphs \cite{Ide,Priebe,Akoglu3,Neil,Rossi,Wang,Palladino} is not applicable to our approach, since the novel vertices we wish to assess lack histories.  

The approach called ``OddBall'' \cite{Akoglu2} identifies anomalous substructures in graphs, in particular, near-cliques and stars, heavy vicinities, and dominant heavy kinks.  These anomalies are dense neighborhoods and ``chatty'' vertices, useful for application to certain kinds of communication, like email and peer-to-peer networks.  Anomalies are spotted in their relation to interesting scaling behaviors that the tested complex networks exhibit with respect to these substructures.  Our login graphs are very small in comparison (typically with only dozens of nodes and edges) and do not exhibit these scaling behaviors; furthermore, the anomalies we seek are unlikely to be particularly chatty since they represent stealthy lateral movement. A similar kind of anomaly is targeted in \cite{Shetty}, in which ``important nodes'' are found using an entropy-based measure; when applied to the Enron email corpus, these vertices tended to be quite central with respect to influence on other vertices.   The work of \cite{Nobel} finds common substructures within a graph as those that are most compressible, analyzed via the MDL;  the least common substructures can be considered anomalous.  
 
Methods that identify outliers in graphs via clustering or community detection include \cite{Moriano,Xu,Chakrabarti}.  Outliers in attributed graphs can be found with the methods of \cite{Gao,Muller}, which apply probabilistic modeling and clustering, respectively, to identify anomalous vertices.   Each of these methods is particularly designed for large networks, with $\geq \mathcal{O}(100)$ vertices, which, unlike our login graphs, are large enough to exhibit rich community structures. In \cite{Lamprakis}, isolated nodes are identified as anomalous, but this application is to the detection of command and control channels which are expected to lack connections to other internal systems.

Methods that attempt to identify anomalous graphs in relation to other graphs proceed by devising a useful way of summarizing the global properties of each graph \cite{Ide,Gaston,Wilson,Berlingerio,Macindoe,Li12,Gamachchi1}, or by quantifying pairwise similarities among graphs \cite{Pincombe,Shoubridge99,Shoubridge02,McWherter,Peabody,Bunke04,Bunke06,Koutra}.  Graph spectra are used for anomaly detection in \cite{Ide} and for classification and clustering in \cite{Wilson}.  A modified notion of graph diameter is analyzed via time series techniques in \cite{Gaston}.  The work of \cite{Macindoe} assigns graphs to points in a three-dimensional space spanned by leadership, bonding, and diversity distance.  The authors compute pairwise distances between graphs and use hierarchical clustering to find groups and outliers.  In \cite{Li12}, a large set of features, including summary statistics of different centrality measures, are used to train a support vector machine to classify graphs.  A similar approach based on one-class learning is taken in \cite{Gamachchi1}, in which graphs are summarized according to measures like vertex and edge count, density, and diameter, and analyzed by an isolation forest to spot outliers.  Global features, including the first several statistical moments of vertex centralities and egonet characteristics, are used to find similarities between graphs in \cite{Berlingerio} via Canberra distance.  

The so-called $\lambda$- (or spectral) distance is developed in \cite{McWherter,Peabody,Bunke04}, which quantifies graph similarity in terms of the Euclidean distance between vectors of eigenvalues of graph adjacency and related matrices (the graph spectra).  In \cite{Pincombe}, a set of graph difference metrics, including maximum common subgraph, edit, modality, diameter, and entropy distances, are analyzed as an ARMA time series to identify anomalous graphs in an evolving sequence.  A notion of node affinity between graph pairs is proposed in \cite{Koutra} and a score is derived for use in outlier detection.  

Finally, the works of \cite{Henderson,Henderson10} describe methods of feature extraction from graphs, with the goal of including local, neighborhood, and global characteristics.  Recursive features are proposed in \cite{Henderson}, in which a particular local measure is aggregated over a vertex's neighbors and assigned to that vertex; in this way, what are formally vertex-level features can summarize neighborhood structure.  In \cite{Henderson10}, a suite of local and community metrics are proposed that characterize the centrality, connectivity, and stability of graph structures.  These works do not propose anomaly detection schemes; however, the work of \cite{Palladino} employs the recursive features of \cite{Henderson} to develop a graph-based intrusion detection system.
 
\section{Login graphs and features}
\subsection{Login graphs}
\begin{figure}[htp]
\centering
\includegraphics[width=2in]{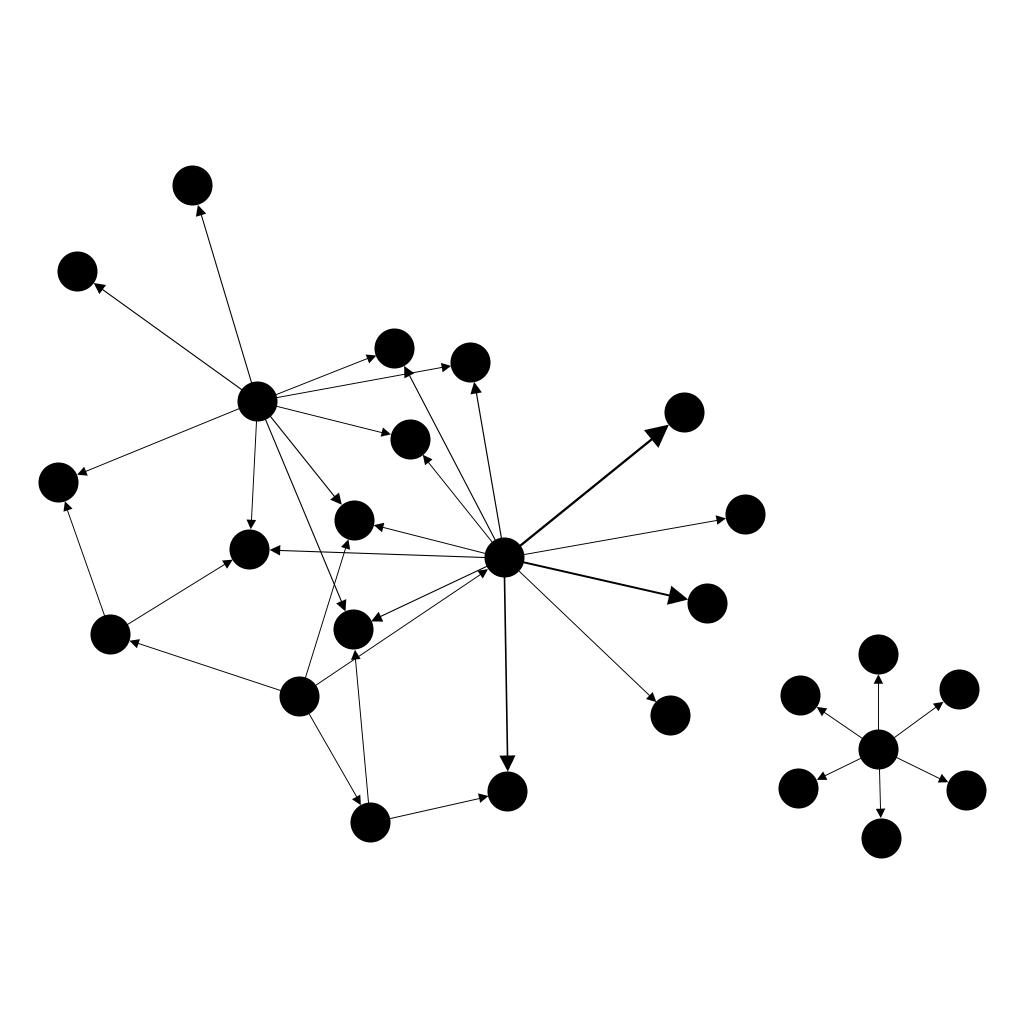}
\caption{\footnotesize{Daily login graph of a user.  Vertices are systems and directed edges indicate logins. Edges are weighted according to number of logins occurring over the 24-hour period.}}
\label{1}
\end{figure}

The basic data structure used in this analysis is the daily {\it login graph} of a particular user, Figure \ref{1}.  It is a weighted, directed graph depicting all the user's remote login activity over a 24-hour period.  We write $G^u_i$ for user $u$'s login graph on day $i$. The vertices are information systems and a directed edge $(u,v)$ between vertices $u$ and $v$ represents successful authentication from system $u$ to system $v$.  The edges are weighted according to the number of logins that occur over the time period covered by the login graph.  The vertex $v_{i,j} \in G^u_i$ corresponds to a unique system and carries two labels: the day, $i$, and a unique identifier like an IP address or host name, $j$.  The user's login history is then the sequence of login graphs $\mathcal{G}^u = \{G^u_1,G^u_2,...,G^u_m\}$ for $m$ days of activity.  Note that the same system, $j$, appearing on two different days $p$ and $q$ is represented by two distinct vertices, $v_{p,j}$ and $v_{q,j}$.  

Login graphs like the one in Figure \ref{1} can be generated using any authentication log that includes the following data elements: user, source and destination systems, and a timestamp of the login.  In this study, we use Windows Event Logs\footnote{In many enterprise environments, Windows Event Log data is aggregated into a SIEM like Splunk and can be easily queried.  If not, client logs can be obtained from individual systems and server logs from Domain Controllers.} queried from a SIEM receiving these logs on a near-continuous basis from clients and servers across the network.  We consider successful logon events (event type 4624) of any login type (both network and interactive).  We omit network logins to Domain Controllers since these records are authentication validations and do not represent actual logins to the Domain Controllers.  

We next decide on the time period over which we wish to model the user's login behavior, which should be long enough such that the model is stable and accurate\footnote{As assessed via a process akin to cross-validation; we describe this process later.}.  In this work, we find that four weeks is a sufficient amount of time to construct reliable models for the great majority of the 78 users considered in this analysis.  

\begin{figure*}[ht]
\centering
\includegraphics[width=5in]{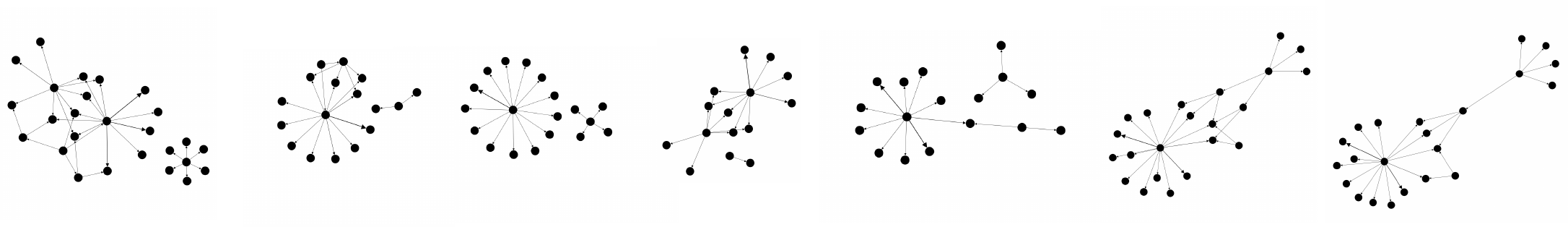}
\caption{\footnotesize{Seven consecutive days of a user's login activity.  There is considerable variability in the number of systems accessed and the topology of the login graph.}}
\label{2}
\end{figure*}

From day to day, a given user's login activity can vary greatly.  In Figure \ref{2}, seven consecutive days of login graphs of the same user are shown: they vary considerably in the number of vertices, edges, and general topology. We will see later on that this variability makes anomaly detection based on global graph characteristics very difficult to apply, motivating the need to quantify graph topology more locally. 

\subsection{Graph measures}
We consider a variety of features that characterize the centrality and local neighborhood topology of individual graph vertices.  We select measures that are potentially useful in characterizing normal login activities or distinguishing malicious login activities.  These include: in degree ($k_{\rm in}$), which is the number of vertices that connect to $v$; out degree ($k_{\rm out}$), the number of vertices that $v$ connects to; in weight ($w_{\rm in}$), the sum of all incoming edges (including multiplicities); out weight ($w_{\rm out}$), the sum of outgoing edges (including multiplicities); left Katz centrality,
\begin{equation}
\label{katz}
e_K(v) = \alpha \sum_s A_{vs}e_K(s) + \beta,
\end{equation}
where $\alpha$ is an attenuation factor, $\beta$ is an initial centrality given to all vertices, and $\mathbf{A}$ is the graph adjacency matrix; local clustering coefficient,
\begin{equation}
c(v) = \frac{2T(v)}{k(v)(k(v)-1)}
\end{equation}
where $T(v)$ is the number of triangles through $v$ and $k(v)$ is its degree; number of edges in $v$'s egonet, that is, the subgraph of $v$'s neighboring nodes,
\begin{equation}
E(v) = k(v) + \sum_{s \in N} k(s) 
\end{equation}
where the sum is taken over all nodes connected to $v$; and eccentricity, $\epsilon$, which is the longest directed path from the vertex.  We also consider a few quantities derived from these base measures, including degree $k = k_{\rm in}+k_{\rm out}$, two kinds of ``reduced'' eccentricity: $\epsilon/(E+1)$ and $\epsilon/(w_{\rm out}+1)$, and two rescaled degree measures: $k_{\rm in}/(w_{\rm in}+1)$ and $k_{\rm out}/(w_{\rm out}+1)$.  Eccentricity is particularly useful for identifying long chains of logins indicative of lateral movement: the ego-reduced eccentricity is sensitive to root vertices that originate long chains and have few other neighbors; the weight-reduced eccentricity is sensitive to login chains of low weight, potentially useful for spotting the stealthy adversary that limits redundant logins between the same systems.  The rescaled degree measures are based on a similar idea: they are sensitive to large in- or out-degree systems with low-weight edges, indicative of an adversary that makes very few redundant logins to many different systems. 

The full collection of graph measures is given in Table \ref{tab1}.
\begin{table}
\begin{center}
\begin{tabular}{|c|c|}
\hline
Index& Measure\\
\hline
0&out degree, $k_{\rm out}$\\
1&out degree (rescaled), $k_{\rm out}/(w_{\rm out}+1)$\\
2&in degree, $k_{\rm in}$\\
3&in degree (rescaled), $k_{\rm in}/(w_{\rm in}+1)$\\
4&local clustering, $c$\\
5&Katz centrality, $e_K$\\
6&ego degree, $E$\\
7&out weight, $w_{\rm out}$\\
8&in weight, $w_{\rm in}$\\
9&degree, $k$\\
10&eccentricity (ego-reduced), $\epsilon/(E+1)$\\
11&eccentricity (weight-reduced), $\epsilon/(w_{\rm out}+1)$\\
12&eccentricity, $\epsilon$\\
\hline
\end{tabular}
\end{center}
\caption{\footnotesize{Graph topology and centrality measures considered in this analysis.  Measures will be referred to by their index later in the paper.}}
\label{tab1}
\end{table}

\section{Vertex Roles and Outliers}
In this work we seek to detect logins to novel systems, which are systems not accessed by the given user over the course of its historical record prior to the day of test.  Let's call the day of test ``day 0''. Writing the vertex corresponding to such a novel system as $v_{0,j^*}$, we wish to compare it to all other vertices in the historical record, $v_{i,k} \in \mathcal{G} = \{G_{-m+1},G_{-m+2},...,G_0\}$ for $-m+1 \leq i \leq 0$ and all $k \leq |G_i|$.   The graphs in this sequence are not time-ordered, since the novel vertices of interest lack histories and so temporal properties cannot be used to identify anomalous behavior.  Instead, we consider all vertices across all login graphs together and attempt to organize them into roles.  The hope is that malicious logins to novel systems will not fit well into any of these roles, and appear anomalous. 

This approach to graph anomaly detection, in which vertices are assigned to roles, has been previously explored \cite{Rossi,Henderson2,SunNMF}.  The key to role assignment is a compressive function that reduces the original feature space of the vertices to lower-dimension, with the idea that the more parsimonious features isolate the most important variations in the data.  The choice of compressive function is arbitrary: we compare performance with non-negative matrix factorization (NMF) and principal component analysis (PCA). Since the compression involves a loss of information, the inverse transformation does not perfectly recover the original feature vectors.   The transformation to the reduced basis set is largely informed by the bulk of normal points in the sample, and so unusual points will not be reconstructed well by the inverse transformation.  This {\it reconstruction error} can be examined for each vertex, and those with large errors can be flagged as anomalous. 

\subsection{Non-negative Matrix Factorization}
Non-negative matrix factorization (NMF) is primarily a dimensionality reduction technique (see \cite{Gillis} for a nice review).  It has found success in a range of applications, including image segmentation and text mining, for its ability to find efficient, sparse representations of high-dimensional feature spaces.  From $n$ data points $x_i \in \mathbb{R}^p$, the matrix $\mathbf{X} \in \mathbb{R}^{n\times p}$ is formed.  The objective of NMF is to find two non-negative matrices $\mathbf{G}$ and $\mathbf{F}$ such that
\begin{equation}
X_{np} \approx G_{nr}F_{rp},
\end{equation}
where $r < \min(p,n)$ is the number of {\it roles}, or the dimensionality of the reduced set of basis elements. Since each data point $x_i$ will generally have non-zero components along each of the $r$ directions, it will belong to a mixture of roles.  The factorization is found by minimizing the Frobenius norm of the error,
\begin{equation}
\label{error}
||\mathbf{X} - \mathbf{G}\mathbf{F}||^2_F = \sum_{i,j} (X - GF)^2_{ij},
\end{equation}
with $\mathbf{G} > 0$ and $\mathbf{F} > 0$. NMF is a non-convex optimization problem, and so a single matrix $\mathbf{X}$ can have many possible factorizations.  For this study, we employ coordinate descent for the optimization and initialize with the singular value decomposition of $\mathbf{X}$ to enforce sparsity of the factorization, that is, to ensure that points tend to map into single roles.  With sparse representations and/or those with fewer roles than the dimension of the original feature space, NMF results in a compression of the original data that necessarily involves the loss of information.  The amount of information lost in transforming a data point $x_i$ is quantified in terms of the transformation's {\it reconstruction error},
\begin{equation}
\label{xerror}
\delta_{x_i} = \sum_{j,k}(X_{ij} - G_{ik}F_{kj})^2.
\end{equation}
Points not well-reconstructed by the inverse transformation will have relatively large errors, and anomalies can be identified by imposing a threshold on this value.  A related approach to outlier detection based on the residuals of the NMF transformation was explored in \cite{Tong}.
 
To apply this approach to our graph vertices, we first evaluate each vertex according to some subset of the measures in Table \ref{tab1}; for example, ${\bf f}(v_{i,j}) = (k_{\rm out}(v_{i,j}),c(v_{i,j}),\epsilon(v_{i,j}))$.  Each of these feature vectors forms a row of the matrix $\mathbf{X}$, indexed by day $i$ and label $j$.  There are $\sum_i |G_i| = n$ vertices in total, each evaluated across $p$ graph measures and so the matrix $\mathbf{X}$ has dimensions $n\times p$. 
 
\subsection{Principal Component Analysis}
Principal component analysis (PCA) is a change of basis to a linearly uncorrelated set of feature vectors, called {\it principal components}. The first component points in the direction of maximum variance, and likewise for subsequent components under the constraint of mutual orthogonality.  Let $\mathbf{X}$ be the $n\times p$ matrix of feature vectors ${\bf f}(v_{i,j}) = {\bf f}_{s}$ of $p$ measures.  We form the covariance matrix
\begin{equation}
\mathbf{\Sigma} = \sum_{s}({\bf f}_{s} - \overline{{\bf f}})({\bf f}_{s} - \overline{{\bf f}})^\intercal.
\end{equation}
The principal components, ${\bf u}_t$, satisfy $\mathbf{\Sigma} {\bf u}_t = \lambda_t {\bf u}_t $, with the $\lambda_t$ the eigenvalues of the covariance matrix.  Since $\dim {\bf u}_t = \dim {\bf f}_{s}$, PCA is in general not a dimensional reduction, but frequently only the first few components are kept as the most relevant features.  Suppose only the first $r$ components are retained: let $\mathbf{\Lambda}$ be the $n\times r$ matrix made from the first $r$ eigenvectors of $\Sigma$.  Then, the transformed features are $\tilde{{\bf X}} = {\bf X}{\mathbf{\Lambda}}$. The inverse transformation is carried out by ${\mathbf{\Lambda}}^\intercal$ thus,
\begin{equation}
\hat{{\bf X}} = \tilde{{\bf X}}{\mathbf{\Lambda}}^\intercal + \overline{{\bf X}}.
\end{equation}
The reconstruction error for a general data point $x_i$ is then taken to be the Euclidean distance between the two vectors,
\begin{equation}
\delta_{x_i}^2 = \sum_j (X_{ij} - \hat{X}_{ij})^2.
\end{equation} 
A related outlier detection scheme based on PCA reconstruction error is given in \cite{Shyu}.

\begin{figure}[ht]
\centering
\includegraphics[width=4in]{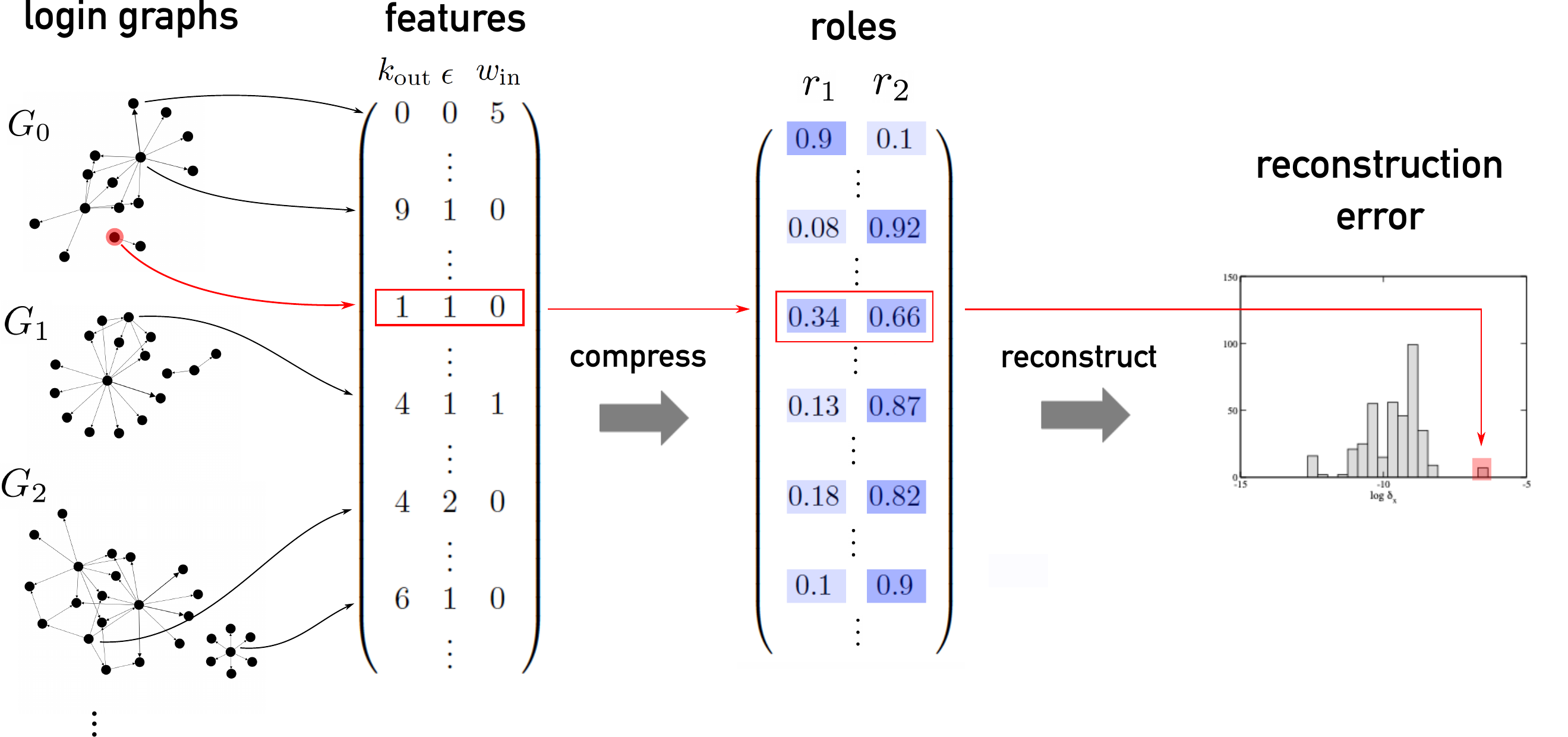}
\caption{\footnotesize{Illustration of the process described in Section 4.  All vertices from each of a user's login graphs are evaluated across a set of graph features and transformed together via a compressive operation (here, NMF or PCA).  Each vertex is assigned a role by this reduction, and vertices which do not fit well into these roles are potential outliers.  The compression is then inverted to recover the original feature representation of each vertex, and the reconstruction error is used to quantify anomalous behavior. The vertex highlighted in red is an example anomaly.}}
\label{nmf_demo}
\end{figure}

An illustration of the above-outlined procedure for general compression function is shown in Figure \ref{nmf_demo}.  

\section{Building User Models}
To profile a user's historical login behavior, we must select a set of graph measures, $\mathbf{m} = \{m_1,m_2,...,m_p\} \in \mathcal{M}$ from Table \ref{tab1} and decide on the dimension of the compressed space, $r$. When using NMF, traditionally the number of roles is determined via information criteria or MDL; for PCA, often only the components explaining some fixed percentage of the variance are selected.  Here, our only guidance is that we want a model that performs well at our task of detecting malicious logins to novel systems. Regardless of the compression transformation used, we will refer to the features in the reduced-dimension space as {\it roles}, $r$.  The selection $\mathbf{m}$, choice of compression function, and number of roles, $r$, constitute a {\it model}. 



\subsection{Computing True and False Positive Rates} 
Each model is assessed in terms of its false positive rate (FPR)---the percentage of benign novel logins it detects as malicious, and its true positive rate (TPR)---the percentage of malicious logins it detects as malicious; we will explain how we arrive at a TPR for an unsupervised model shortly. 

To establish the FPR of a given model with some $\mathbf{m}$ and $r$, we generate a login graph for each day in the user's historical record.  We then shuffle the set of graphs and split it into two subsets: a larger one of size $N$ (here 80\% of the graphs) and a smaller one.  The smaller one is set aside---it is not used.  Assuming that the login graphs are not correlated across days, the larger set can be thought of as one possible three week history. The matrix $\mathbf{X}$ is formed from the vertices in the larger set, compression is applied, and vertices with reconstruction errors in the top $\alpha \%$ are considered outliers.  The FPR is then computed as the ratio $n_{o,{\rm new}}/n_{\rm new}$, where $n_{o,{\rm new}}$ the number of novel systems that are outliers and $n_{\rm new}$ the number of novel systems (systems with only a single access over the three week history).  This process is repeated \texttt{iters} times (here, \texttt{iters} = 25)\footnote{This number is chosen by trial-and-error; \texttt{iters} is increased until the standard error in FPR across iterations begins to level-out.}: the full set of login graphs is shuffled, split, compression is performed, and outliers are found.  The reason for performing several iterations is to test the generalizability of the model, with each iteration comprising a different possible 3 week history.  The model's final false positive rate is the average over all iterations, $\overline{{\rm FPR}}$.  The FPR testing process described above is summarized as the algorithm in Figure 4. 

\begin{algorithm}
\label{alg1}
\begin{algorithmic}
\REQUIRE{$r$, $\mathbf{m}$, $\alpha$, set of login graphs}
\STATE Initialize array FPR
\FOR{$j=1$ to \texttt{iters}} 
\STATE $\mathcal{G}$ $\leftarrow$ randomize login graph subset of size $N$
\STATE $\mathbf{X}$ $\leftarrow$ compute graph measures $\mathbf{m}$ of all nodes in $\mathbf{g}$
\STATE apply compression with $r$ roles to $\mathbf{X}$
\STATE compute reconstruction errors
\STATE FPR$[j]$ $\leftarrow$ compute $n_{o,{\rm new}}/n_{\rm new}$ at threshold $\alpha$ 
\ENDFOR
\STATE $\overline{{\rm FPR}} \leftarrow$ mean$_j$ FPR
\RETURN $\overline{{\rm FPR}}$ 
\end{algorithmic}
\caption{\footnotesize{Algorithm to compute false positive rate of a model with measure set $\mathbf{m}$, number of roles, $r$, and significance level, $\alpha$. In this study, \texttt{iters} = 25.}}
\end{algorithm}

Computing the true positive rate for a given model is more challenging.  Anomaly detection systems based on unsupervised learning lack labeled anomalous data points.  This makes model validation difficult: we can tune our model to achieve a low false positive rate on the presumed normal data, but how do we know whether it will perform well against the kinds of malicious activities we'd like it to detect?  To address this problem, we generate a small collection of prototype adversarial login graphs and use them to validate the model's true positive rate.  It is important to emphasize that at no point is the model {\it trained} on these prototypes---it is an unsupervised method applied to unlabeled data.  The validation set therefore does not need to be exhaustive, but includes only prototype patterns indicative of adversarial lateral movement.  

To generate these patterns we apply a few simple rules that describe the generic behavior of adversarial lateral movement.  The first rule is that the adversary never ``doubles back'', that is, if they login to system $v$ from system $u$, they will never login to system $u$ from $v$.  This is because access to the originating system, in this case $u$, is assumed to be preserved throughout the course of the intrusion.  Second, the adversary never accesses a system $u$ from more than one other system, since a single access should be sufficient to establish a foothold on that system and multiple logins from distinct systems might arouse suspicion.  These rules imply that each graph has a single root vertex with $k_{\rm in} = 0$; this is the system from which the adversary originates further access.

\begin{figure}[ht]
\centering
\includegraphics[width=4in]{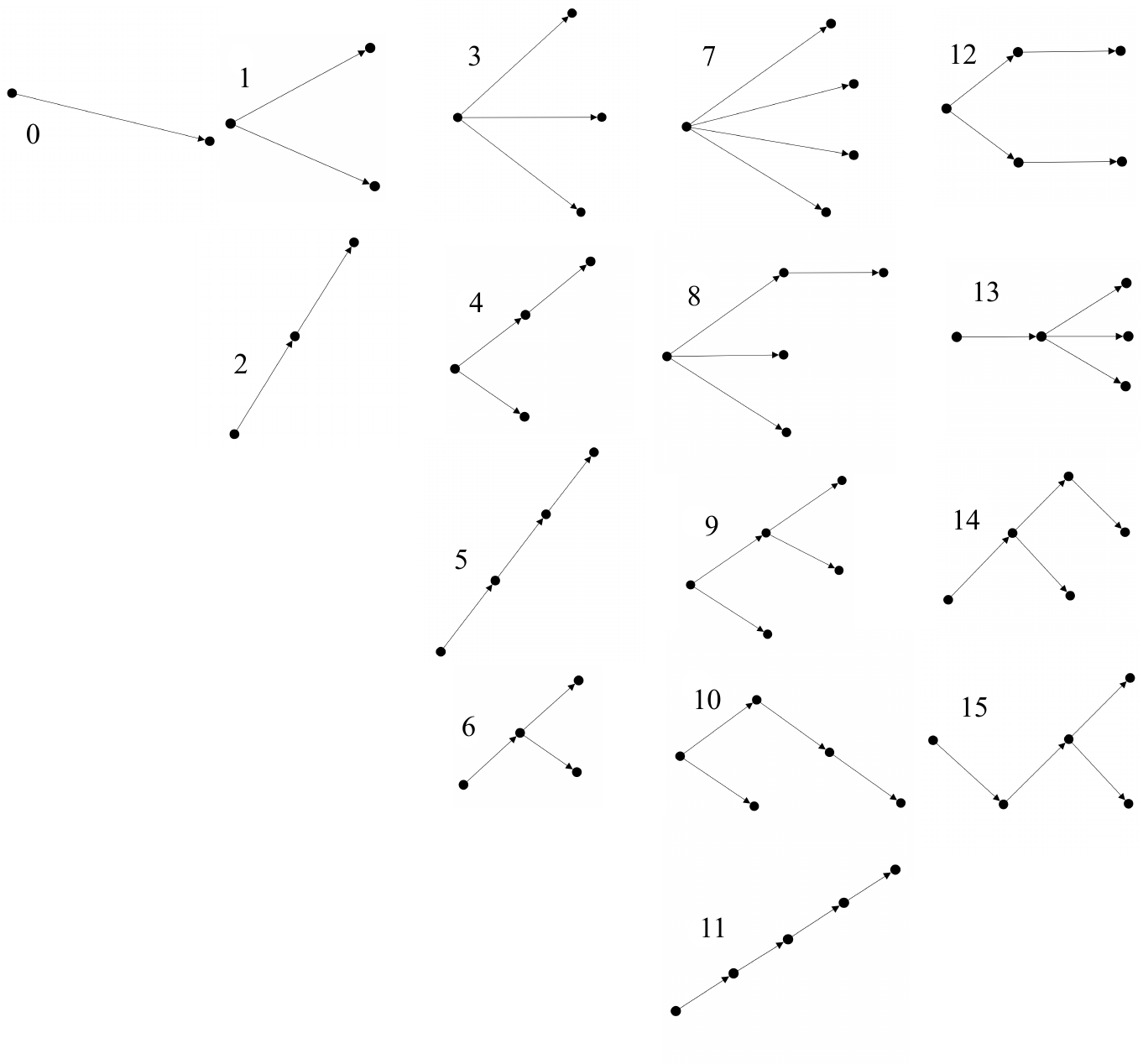}
\caption{\footnotesize{The complete set of 16 possible adversarial graphs including between 2 and 5 vertices.}}
\label{attack_graphs}
\end{figure}
In this study, we allow the number of systems in a graph to vary between two and five. This corresponds to an expected rate of compromise of up to five systems in 24 hours\footnote{We stop at five systems for feasibility; as we shall see, our method works best against larger adversarial graphs and so should perform equally well against graphs with $n > 5$.}. It is then possible to enumerate all possible graphs with $n\leq 5$ that obey the above two rules: there are 16 possible graphs (distinct up to isomorphism), shown in  Figure \ref{attack_graphs}.  For example, there is one possible graph with two systems, namely the edge $(u,v)$.  There are two possible graphs with three systems, namely $\{(u,v),(v,w)\}$ and $\{(u,v),(u,w)\}$.  

\begin{figure}[ht]
\centering
\includegraphics[width=4in]{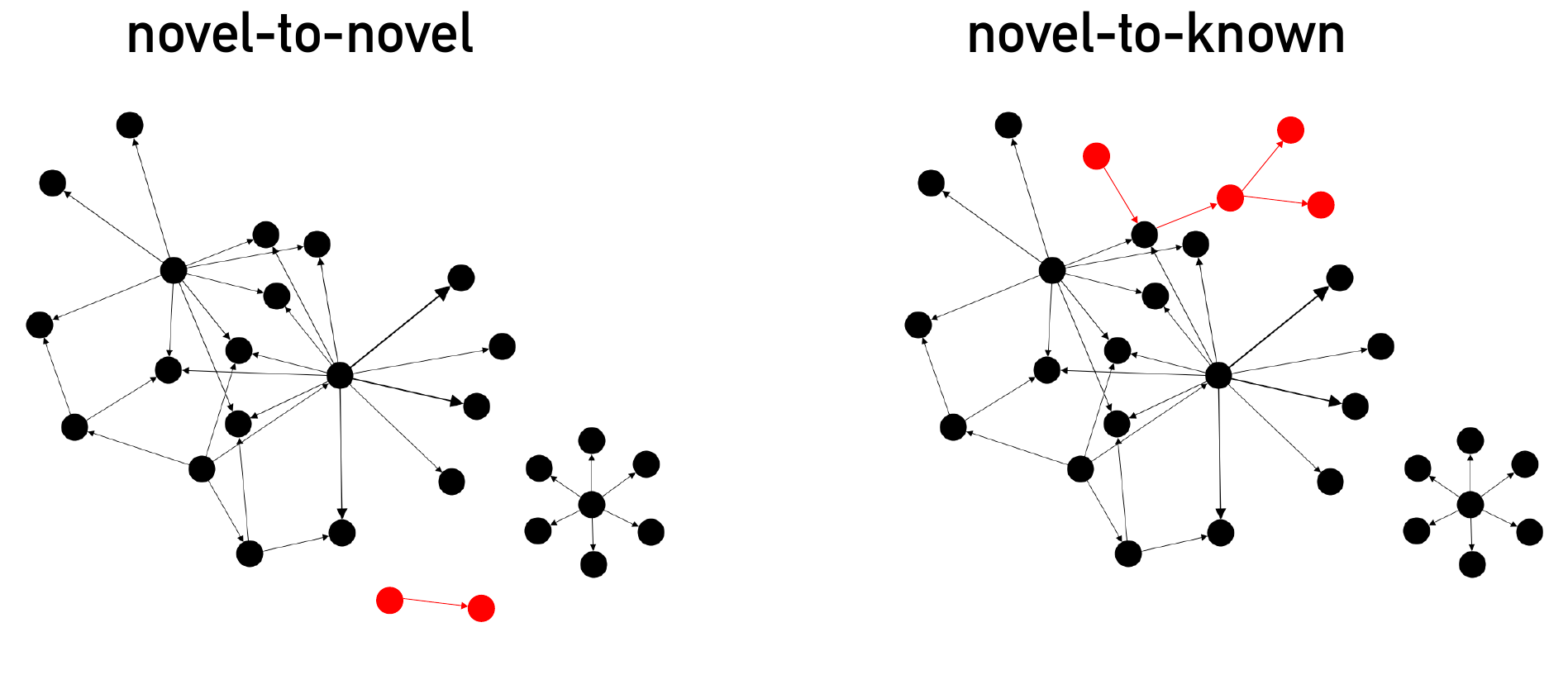}
\caption{\footnotesize{Example login graphs showing each of the two cases considered: novel-to-novel logins, which are separate components, and novel-to-known logins, which attach to random login graph vertices. Adversarial logins are shown in red.}}
\label{sample_adv}
\end{figure}

To measure a model's TPR, we proceed as we did for the FPR determination by repeatedly shuffling and splitting the set of login graphs into a larger set and a smaller set for 25 iterations.  The larger set (again 80\% of graphs) is used for validation by embedding an adversarial graph into a different randomly selected login graph in each iteration.  For the case in which all adversarial logins are novel, the adversarial graph is its own separate component in the parent login graph.  If, on the other hand, the adversary is assumed to traverse known systems as well as novel, the adversarial graph is ``attached'' to the parent graph by replacing one of the adversarial graph's vertices with a randomly selected known vertex from the parent graph.  We will study both of these cases, hereafter referred to as novel-to-novel and novel-to-known\footnote{Despite the choice of name, this case includes access from novel to known as well as known to novel systems.}, respectively, in the Results section. Sample login graphs for each case are shown in Figure \ref{sample_adv}.

If any of the adversarial graph's vertices are detected as outliers, that iteration is considered a true positive.  The TPR is then the number of iterations with detections divided by the total number of iterations (here, \texttt{iters} = 25).  The TPR is computed in terms of iterations rather than vertices (like FPR) since the detection of a single login is sufficient to discover the adversary on that day (iteration).   By including the adversarial graph in a different randomly selected login graph each iteration, we are testing the model's ability to detect that type of adversarial login over a range of possible three-week histories.  A separate set of iterations is performed for each adversarial graph type, $i$, and for each type the TPRs are averaged over the iterations, $\overline{{\rm TPR}}_i$. This algorithm is provided in Figure 6. 

\begin{algorithm}
\begin{algorithmic}
\REQUIRE{$r$, $\mathbf{m}$, $\alpha$, set of login graphs}
\FOR{$i=1$ to \texttt{num\_graphs}}     
\STATE Initialize array TPR 
\FOR{$j=1$ to \texttt{iters}} 
\STATE $\mathcal{G}$ $\leftarrow$ randomize login graph subset of size $N$
\STATE $s$ $\leftarrow$ random integer in range $(1,N)$ 
\STATE inject adversarial graph into login graph $\mathbf{g}[s]$ 
\STATE $\mathbf{X}$ $\leftarrow$ compute graph measures $\mathbf{m}$ of all nodes in $\mathbf{g}$
\STATE apply compression with $r$ roles to $\mathbf{X}$
\STATE compute reconstruction errors
\IF{detection at threshold $\alpha$}
\STATE TPR$[i][j]$ $\leftarrow$ 1 
\ENDIF
\ENDFOR
\STATE $\overline{{\rm TPR}}[i]\leftarrow$ mean$_j$ TPR for graph $i$
\ENDFOR
\RETURN $\overline{{\rm TPR}}[i]$ 
\end{algorithmic}
\label{alg2}
\caption{\footnotesize{Algorithm to compute true positive rate of a model with measure set $\mathbf{m}$, number of roles, $r$, and significance level, $\alpha$. In this study, \texttt{iters} = 25 and \texttt{num\_graphs} = 16.}}
\end{algorithm}

\subsection{Finding well-performing models}
With an understanding of how to measure the FPR and TPR of a prospective model, we next discuss how to find well-performing models.  For each user, we build an NMF-based and PCA-based model for comparison. We first constructed one model for each user with the maximum number of measures (all 13 of them) included in $\mathbf{m}$ and with $r$ set to the number of roles that minimizes the Akaike information criterion for the NMF models, and with $r$ set to the number of components comprising 99\% of the variance for the PCA models.  These models generally performed very poorly, with true positive rates below 25\% for most adversarial graph types for most users. Considering instead the other extreme, we next explored models with the minimum number of measures included in $\mathbf{m}$: NMF and PCA require $|\mathbf{m}| \geq 2$.  With $|\mathbf{m}| = 2$ there are $\binom{13}{2}= 78$ different models for each user; setting $r=1$ and fixing\footnote{After comparing across models with fixed $\alpha$, the receiver operating characteristic curve can be obtained for those of interest by varying $\alpha$.} the significance level to $\alpha = 0.05$, we evaluated $\overline{\rm FPR}$ and $\overline{\rm TPR}_i$ for all $i$ for each model for each user.  We repeated this process for all models with $|\mathbf{m}| = 3$, for which there are $\binom{13}{3}= 286$ different models for each user.  Searching over models with $|\mathbf{m}| > 3$ quickly becomes computationally infeasible; luckily, we find good models with $|\mathbf{m}| = 2$ and 3 and so we stopped there. 

\begin{figure}[h]
\centering
\includegraphics[width=3.5in]{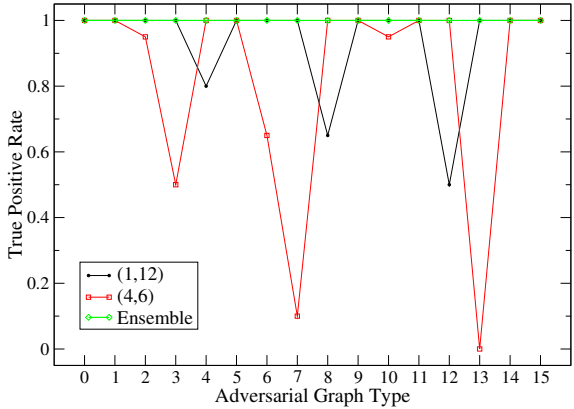}
\caption{\footnotesize{True positive rate vs. adversarial graph type for two models, $(1,12)$ (black) and $(4,6)$ (red), together with the ensemble formed by the logical OR of their detections (green).}}
\label{ens_ex}
\end{figure}

Testing first against novel-to-novel adversarial logins, we found that, in general, no single model performed strongly across the full range of 16 adversarial graph types for any user.  However, with so many models to work with for each user (78 2-dimensional and 286 3-dimensional), it becomes possible to combine different models into an {\it ensemble} that performs well across all adversarial graphs.  For example, for one user under NMF, by taking the logical OR of the $(1,12) = (k_{\rm out}/(w_{\rm out}+1),\epsilon)$ and $(4,6) = (c,E)$ models, neither of which performs consistently well across all adversarial graph types, we obtain an accurate ensemble model, Figure \ref{ens_ex}.  Of course, while ensembles offer generally better TPR rates across different adversarial graph types, we expect the $\overline{{\rm FPR}}$ of the individual models to be generally additive.  

\begin{figure}[h]
\centering
\includegraphics[width=3.5in]{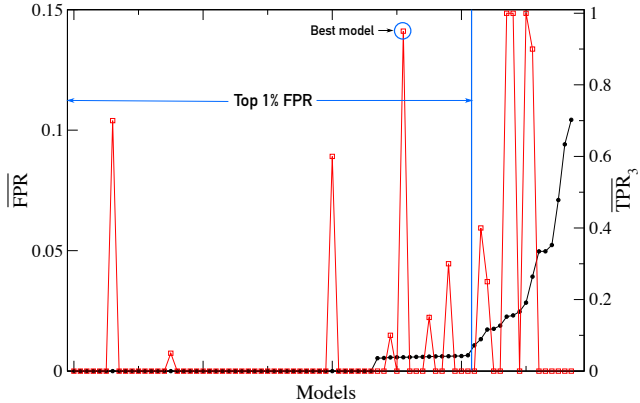}
\caption{\footnotesize{Example of how ensemble models are selected.  Figure shows $\overline{{\rm FPR}}$ (black) and $\overline{\rm TPR}_3$ (red) of each of the 78 two-parameter models for adversarial graph type 3.  The model with the best $\overline{\rm TPR}_i$ within the top 1\% of $\overline{{\rm FPR}}$ is selected.}}
\label{ex}
\end{figure}

To build ensembles with low FPR, we rank each model within each adversarial graph type $i$ according to $\overline{{\rm FPR}}$, lowest first, keeping only the top 1\% of models.  Then, we choose the model with highest $\overline{\rm TPR}_i$ within this subset for each $i$.  This process is shown for all $|\mathbf{m}| = 2$ models for a particular user for adversarial graph type 3 in Figure \ref{ex}.  Overall, for NMF we had best performance among models with $|\mathbf{m}| = 2$, and for PCA $|\mathbf{m}| = 3$.

\section{Results}
We now show how the ensembles constructed in the previous section performed for each user using NMF and PCA.  
\begin{figure}[ht]
\centering
\includegraphics[width=3.5in]{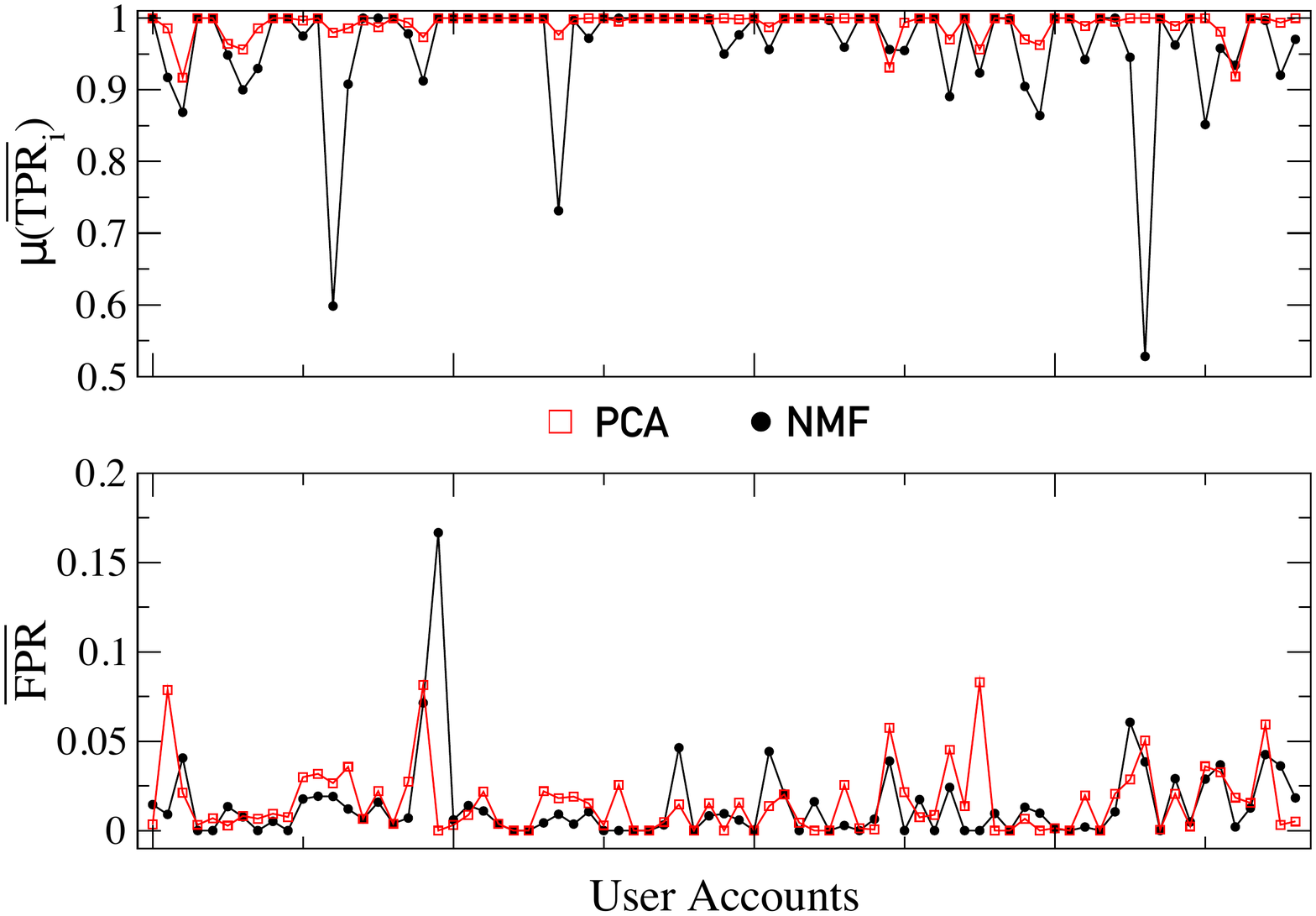}
\caption{\footnotesize{False positive rates ($\overline{\rm FPR}$, solid lines) and average true positive rates ($\mu(\overline{\rm TPR}_i)$, dashed lines) for each user.  Black denotes NMF-based models and red denotes PCA-based models.}}
\label{nmf_v_pca}
\end{figure}
\subsection{Test Users}
We develop ensembles for 78 privileged user accounts using real login data from an operational enterprise network spanning four weeks.  The original group of 92 users had 14 accounts with an insufficient amount of login activity for model construction and so were excluded.  As we explained earlier, model testing requires a 80-20 split of login graphs and so each user must have at least five days of login activity (and, hence, five login graphs).

These privileged accounts fall into five groups based on breadth of access and level of privilege: Domain Administrators, with the highest privilege level, conduct all Domain-related functions; Desktop administrators, with local Administrator rights on most user workstations, are able to manage software deployments and some Domain-related functions; Help Desk administrators, who perform basic account actions like password resets; Server administrators, with local Administrator rights on a variety of Windows file and account servers; and Data administrators, with access to non-Windows, storage, and database systems.  Most accounts were fairly active, with an average of 19 days of login activity over the four week period.  The data used in this analysis are available at \url{https://github.com/bapowellphys/malicious_logins}.
\begin{figure}[ht]
\centering
\includegraphics[width=4.5in]{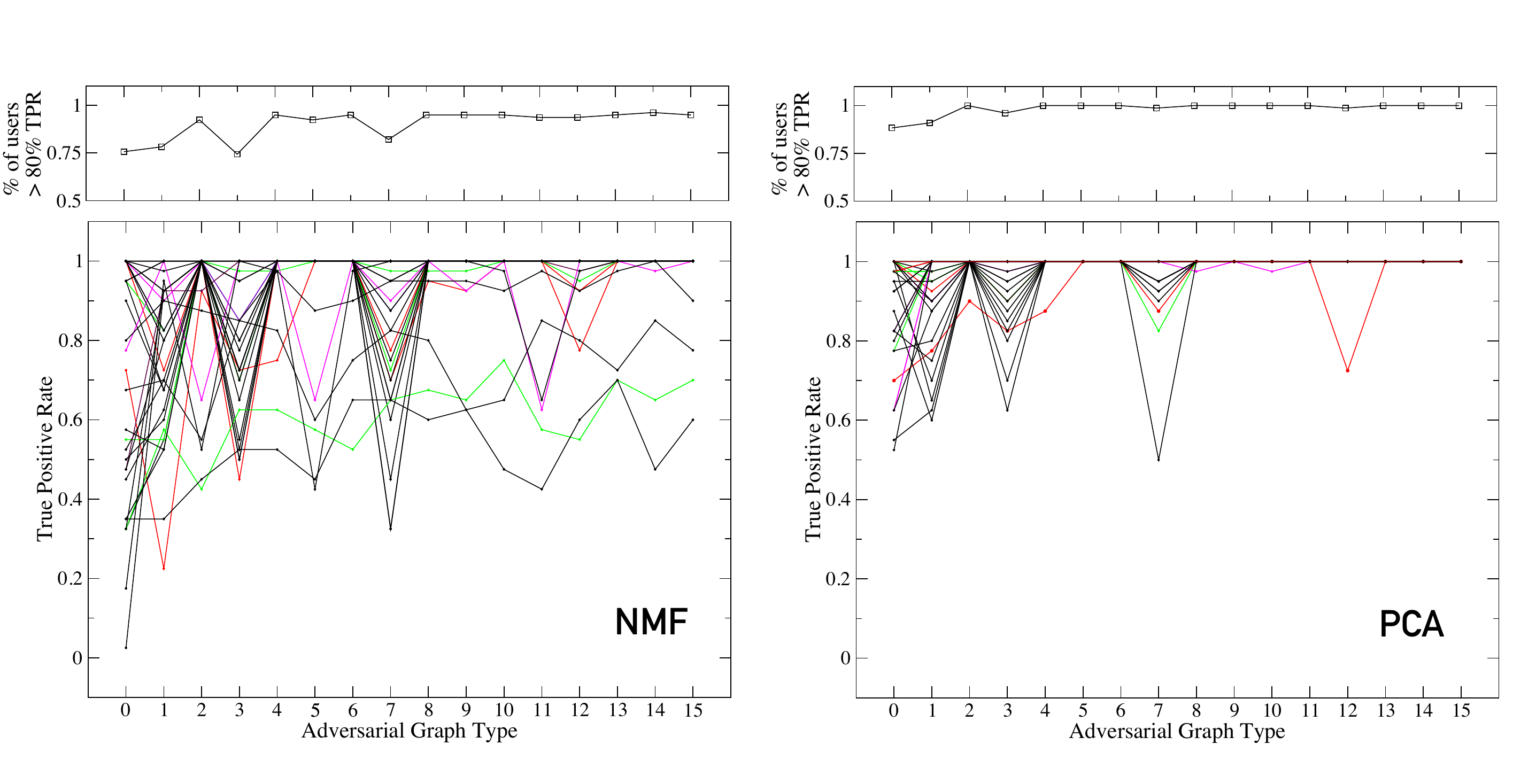}
\caption{\footnotesize{True positive rates ($\overline{\rm TPR}_i$) for graph type $i$ for each user (bottom plots) color-coded according to type of user: Domain Admin (green), Data Admin (magenta), Helpdesk (blue), Server Admin (red), and Desktop Admin (black). The top plots show the percentage of users with a $\overline{\rm TPR}_i$ greater than 80\% for the graph type $i$. Results for NMF are on the left and PCA on the right.}}
\label{tpr_by_graph}
\end{figure}
\subsection{Model performance against adversarial logins}
We first present results of ensembles developed for the novel-to-novel case.  User models are scored by $\overline{{\rm FPR}}$ and $\mu(\overline{\rm TPR}_i)$, the average true positive rate for graph type $i$, averaged over all graph types. Results are shown in Figure \ref{nmf_v_pca}.  The average $\overline{{\rm FPR}}$ for NMF is 1.4\% and PCA is 1.6\%; the average $\mu(\overline{\rm TPR}_i)$ is 96\% and 98\%, respectively.  The NMF- and PCA-based models are competitive, though with $|\mathbf{m}| = 3$, it takes longer to search the model space for PCA.  But, NMF has greater computational complexity than PCA and so the latter can be performed more quickly on a typical computer system.  These two factors make NMF- and PCA-based model building roughly equivalent in terms of computational resources.  

Looking more closely at how each user does against each graph type, we plot $\overline{\rm TPR}_i$ vs $i$ for NFM and PCA in Figure \ref{tpr_by_graph}.  The color-coding reflects the type of account: Domain Admin (green), Data Admin (magenta), Helpdesk (blue), Server Admin (red), and Desktop Admin (black).  The top plots show the percentage of users whose models have $\overline{\rm TPR}_i > 80\%$ for that graph type.  NMF is worse, with more users with models that perform poorly against more graph types.  Notable from these plots is that certain adversarial graphs, like $i=0,1,3$ and 7 tend to give many users problems for both NMF and PCA. Looking at Figure \ref{attack_graphs}, these graph types are all ``fan-like'', with $k_{\rm out} = 0,2,3$ and 4, suggesting that this kind of authentication pattern, in which single logins are made from a central system, is not unusual behavior for many accounts.  

\begin{figure}[t]
\centering
\includegraphics[width=4in]{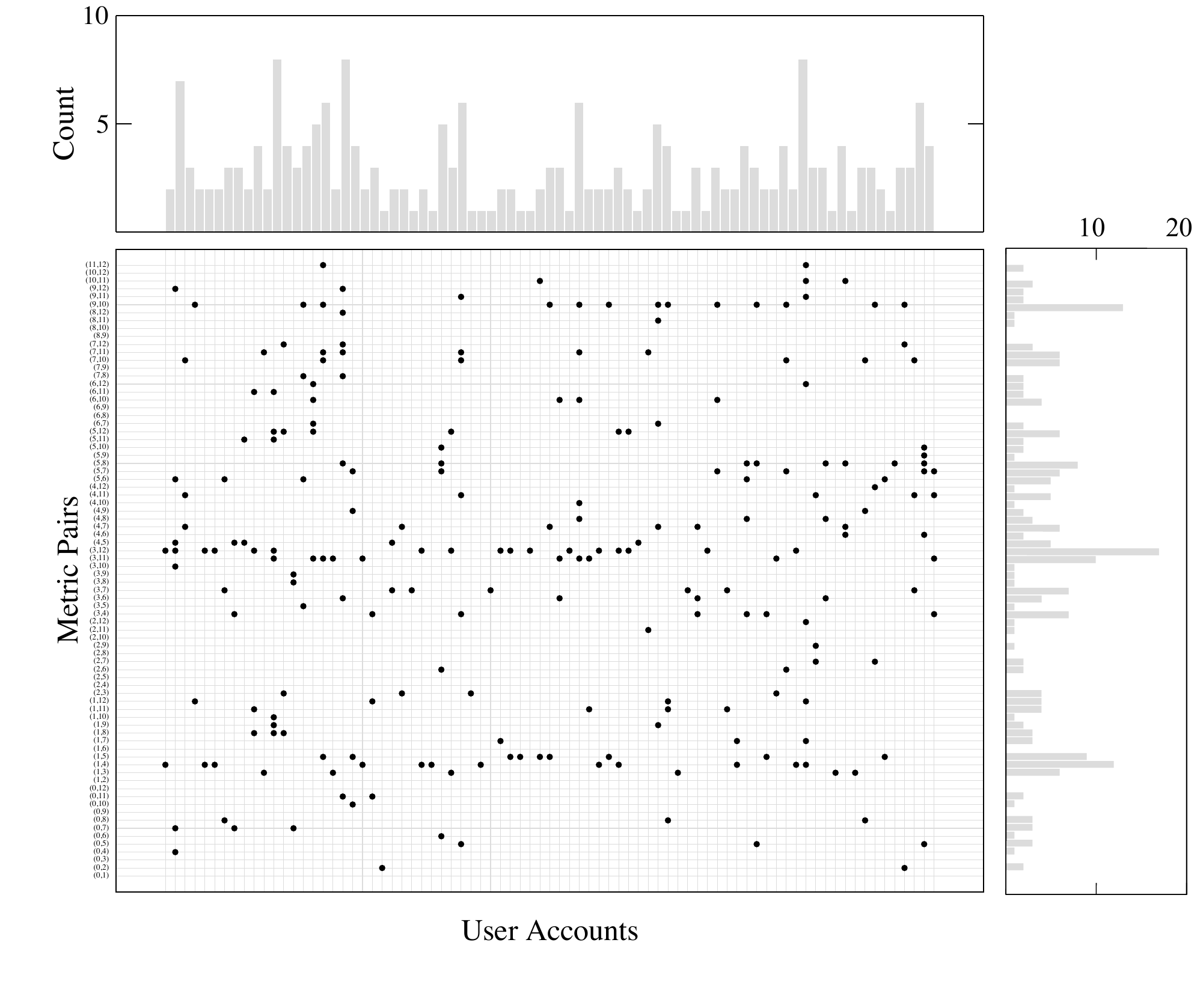}
\caption{\footnotesize{Metric pairs included in the NMF ensemble of each user.  Dots indicate that a measure pair is included in the ensemble of the user on the axis vertically below the dot.  Bar charts are histograms showing the number of pairs in each user's ensemble (top) and the number of times a particular pair is used in an ensemble (side).}}
\label{models}
\end{figure}

To get a sense of what these ensembles look like, we present the best NMF ensembles of each user in terms of its component two-parameter models in Figure \ref{models}.
We note that virtually all measure pairs are used in at least one ensemble, though there is a clear preference for certain pairs, like $(1,4) = (k_{\rm out}/(w_{\rm out}-1),c)$, $(9,10) = (k,\epsilon/(E+1))$, and $(3,12) = (k_{\rm in}/(w_{\rm in}+1),\epsilon)$.  Also, some users get by with only a single two-parameter model whereas others have ensembles of up to eight component models.

We next present results of models developed for the novel-to-known case following the same procedure described in Section 5 for novel-to-novel models; the only difference is that adversarial graphs are now ``attached'' to the parent login graph as described in Section 5.1 and shown in Figure \ref{sample_adv}.  Overall, these kinds of logins are more challenging to detect.  Because the adversarial graphs are no longer separate components in their parent login graphs, the feature vectors of its vertices depend on the nature of the login graph into which it is embedded (in particular, measures that characterize the local neighborhood of the graph like ego degree, Katz centrality, and eccentricity will vary by login graph). For the novel-to-known case, validating the model over multiple iterations thus does more than just test the generalizability of the model given alternate histories---it assesses the model's ability to detect a particular adversarial graph in a variety of different daily login scenarios.   
\begin{figure}[ht]
\centering
\includegraphics[width=3.5in]{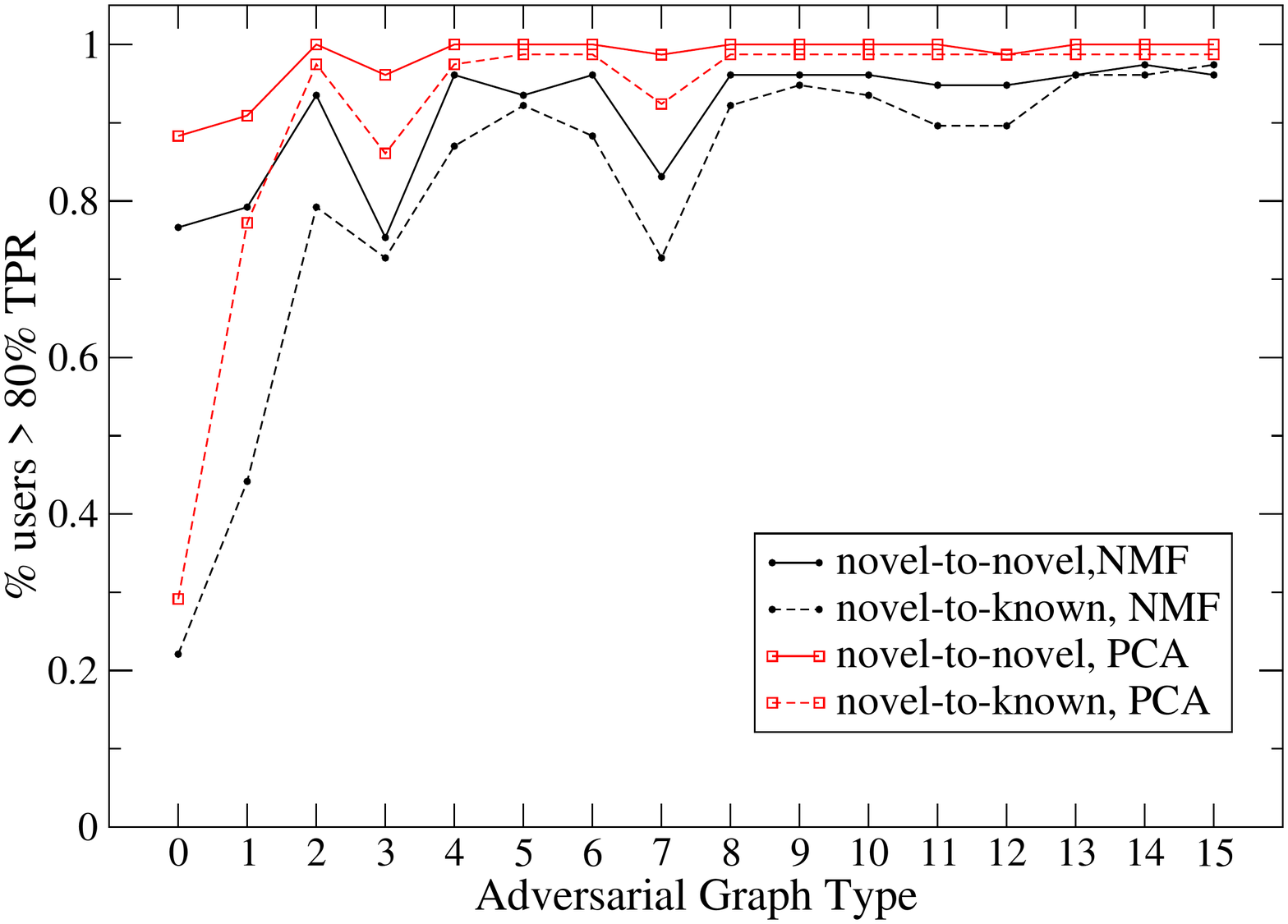}
\caption{\footnotesize{Comparison of NMF and PCA reconstruction error-based models against novel-to-novel and novel-to-known logins, showing percentage of users with $\overline{TPR}_i > 80\%$ as a function of graph type, $i$.}}
\label{1v4}
\end{figure}

NMF models are degraded: the average $\overline{{\rm FPR}}$ is 2.5\% and average $\mu(\overline{{\rm TPR}}_i)$ drops to 91\% (compare 1.4\% and 96\% for novel-to-novel).  PCA models, on the other hand, largely retain their sensitivity, with an average $\overline{{\rm FPR}}$ of 2.2\% and an average $\mu(\overline{{\rm TPR}}_i)$ of 96\% (compare 1.6\% and 98\% for novel-to-novel).  The average performance of these ensembles against the different adversarial graph types are compared against those of the novel-to-novel case in Figure \ref{1v4}.   Remarkably, the novel-to-known PCA-based model out-performs the novel-to-novel NMF-based model, though only with respect to detections. Considering the entirety of our analysis, PCA is the preferred reduction transformation.

The ensembles developed for the novel-to-known case are generally larger than those constructed for the novel-to-novel case.  This is likely because the number of adversarial graphs is effectively much larger than 16 since they depend on how they are embedded into the parent login graph: the resulting variability is hard for ensembles with fewer degrees of freedom ({\it i.e.} ensembles composed of fewer models) to represent.  

While, for the purpose of exposition, we have constructed separate and different ensembles for each user for the novel-to-novel and the novel-to-known cases, in reality each user should have a single ensemble.  This could have been done from the beginning by computing $\overline{{\rm TPR}}_i$ for each adversarial graph for each use-case.  The resulting ensemble would be the union of the two separate ensembles developed here for each user.  We therefore expect the $\overline{{\rm TPR}}_i$ rates of the composite ensemble to be at least as good as those of the two component ensembles; it is possible, however,  that the $\overline{{\rm FPR}}$ degrades as a result of adding more models.  For our test users, we found this was not the case, with $\overline{{\rm FPR}}$ either remaining the same or degrading only fractionally for each user. 

Lastly, we note that adversarial graphs in which more than a single vertex is replaced by a known system are also possible, corresponding to the case where the adversary authenticates to more than one known system in a 24-hour period.  We don't test these scenarios explicitly but we expect that adding known systems will further degrade performance of the ensembles.  We can infer this by observing, for example, that adding two known systems to the type 1 graph of Figure \ref{attack_graphs} will include the type 0 graph with one known system, as well as the new graph with a novel system authenticating to two known systems.  We might therefore expect performance against type 1 graphs with two known systems to be on a par with performance against the type 0 graph with one known system, and, in general, performance against type $n$ graphs with two known systems should resemble performance against type $n-1$ graphs with one known system. 
\section{Comparison With Other Methods}
In this section, we compare our approach with others from the literature.  We are particularly interested in comparing with other graph-based methods of lateral movement detection, but we also consider alternative features and anomaly detection schemes from other applications.   As a method of detecting compromise via user login graphs, the work of \cite{Kent} is most relevant to ours and we explore it first.
\subsection{Anomalies in aggregated login graphs}
As described earlier, \cite{Kent} builds login graphs of a set of Administrative users over the course of a one month red team exercise at LANL.  Logistic regression is then used as a supervised classifier to detect users that have been compromised by the red team over this time.  In contrast to our method, each user has a single {\it aggregated} login graph spanning several weeks rather than a sequence of daily graphs. The method is aimed at discovering compromised users, not malicious login events, and so anomaly detection is done at the level of the user's aggregated login graph, in comparison to other user's login behaviors.  To compare our methods, we aggregate daily login graphs for each user into a single login graph, $G^u$, spanning the whole of the historical period.  It is defined as the union of vertex and edge sets of each daily graph, $G^u = \bigcup_{i} G^u_i$ for all $G_i^u \in \mathcal{G}^u$.  In our analysis, the same system in two different login graphs was considered two distinct vertices, labeled by day, $i$, and unique identifier, $j$.  To create aggregated graphs, however, we ignore the day label $i$ and consider only the identifier $j$ as discriminating.  This is how the authentication graphs of \cite{Kent} were constructed.

Each user's login graph is evaluated across five global measures in \cite{Kent}: number of vertices, number of edges, time-constrained diameter (which is the maximum-length time-ordered path in the graph), the entropy of the histogram of lengths of time-constrained paths, and the entropy of the histogram of vertex types in the graph.  Vertex types include out-stars (vertices with $k_{\rm out} > k_{\rm in}$), in-stars (with $k_{\rm in} > k_{\rm out}$), isolated vertices (with $k = 0$), time-constrained transit vertices (those with all predecessor vertex connections occurring prior to all successor vertex connections), and pseudo-leaf vertices (all those not falling into the aforementioned categories).  The entropy of histogram measure is simply $S = p_i \log p_i$ where $p_i$ is the relative frequency of bin $i$.  See \cite{Kent} for more details on these definitions.

The primary challenge with the approach of \cite{Kent} is the use of supervised learning, which presupposes an adequate supply of generic intrusion data, which most organizations lack.  While we simulate adversarial graphs in our analysis, these are prototypes of daily activity for which we have fairly generic and reliable models.  Simulating a one-month long campaign is almost guaranteed to be of very low fidelity, and so we forego a comparison of our methods against adversarial activity.  We instead apply one-class learning to our aggregated user login graphs using the five features of \cite{Kent} and compare false positive rates.  

We train a one-class support vector machine (OCSVM) on our 79 aggregated user login graphs using 5-fold cross-validation and find significant variability in FPR across folds.  We find a mean FPR of 0.59 with a standard deviation of 0.36.  This indicates that our group of administrative users exhibits great variability with respect to these five global measures, even when behaviors are aggregated over several weeks.  We conclude, for our test users at least, that this approach to malicious login detection fails.  
\subsection{Substructure anomalies}
In addition to vertex outliers, malicious logins could give rise to substructure anomalies.  A series of logins among novel systems appears in a login graph as a separate component which could be structurally distinct from other groups of vertices in the login graph.  The {\small SUBDUE} algorithm of \cite{Nobel} attempts to identify anomalous substructures, defined as a connected subgraph, by recursively building up subgraphs, compressing them, and evaluating the resulting graphs according to their minimum description lengths.  Specifically, the method minimizes the quantity $L(G|S) + L(S)$, where $L(G|S)$ is the minimum length of graph $G$ after compressing substructure $S$. The least-compressible structures can be deemed anomalous.  In \cite{Eberle07}, the graph-based anomaly detection (GBAD) algorithm was proposed as a customization of {\small SUBDUE}, capable of detecting three kinds of anomaly: insertions, deletions, and substructure modifications.  GBAD was applied to the problem of lateral movement in \cite{Eberle09,Eberle10}.

\begin{figure}[ht]
\centering
\includegraphics[width=4in]{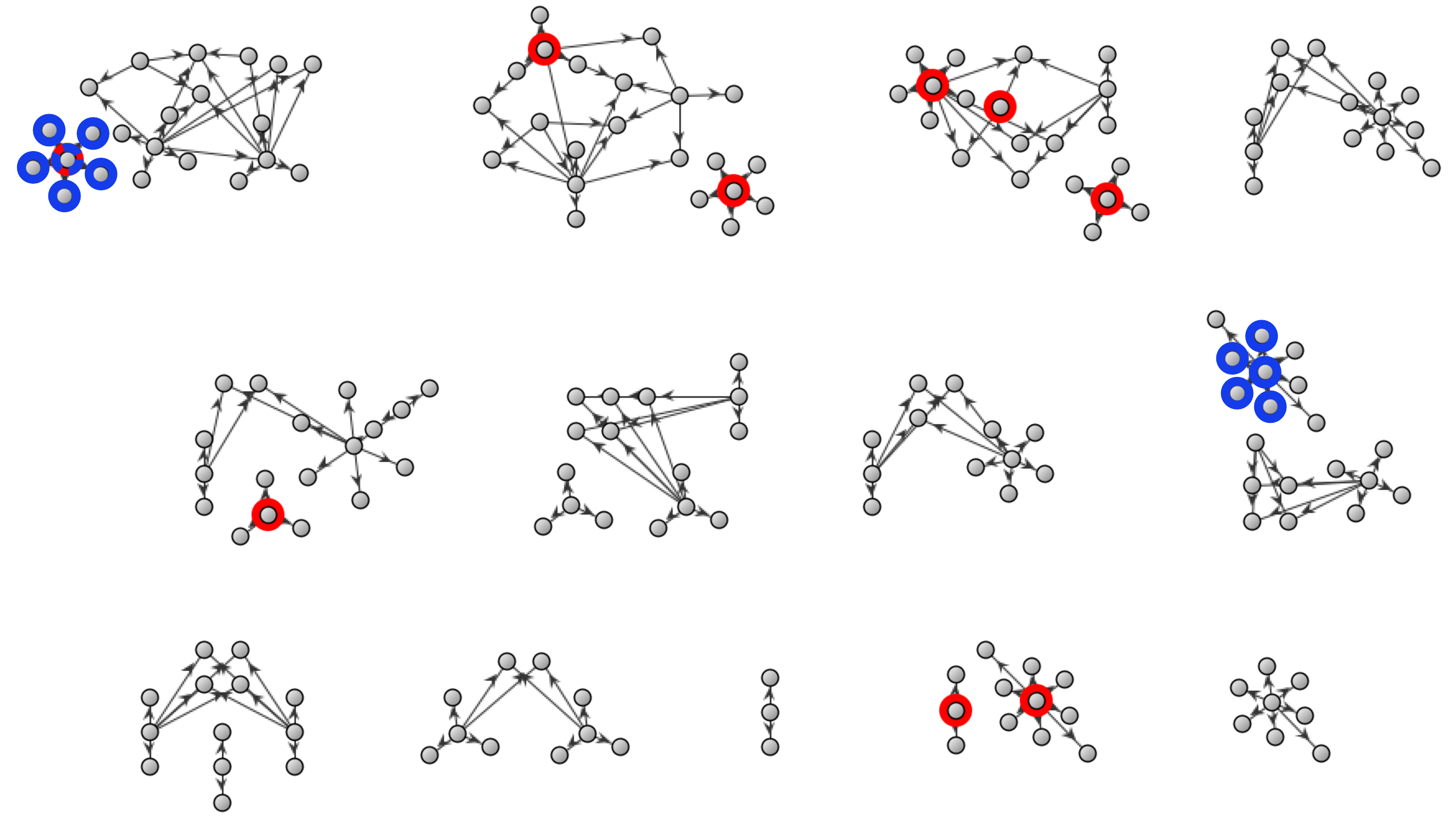}
\caption{\footnotesize{Thirteen normal login graphs for one of the Domain Administrators in our test sample.  Vertices marked in red were detected as malicious by our method, and those in blue by GBAD.  Note that these anomalies are only false positives if the detected systems are novel.}}
\label{comp_anom}
\end{figure}

We apply this approach to each user's login sequence $\mathcal{G}^u$, where substructures are mined across the entire sequence, not within single login graphs for which we expect most subgraphs to be novel given their small size and simplicity.  As an example, consider the complete set of login graphs for one of the Domain Administrators in our test user group, Figure \ref{comp_anom}.  There are 13 login graphs summarizing 13 days of normal login activity: red vertices are those considered anomalous by our method, and blue are those considered anomalous using GBAD.  This example gives a general idea of the performance of GBAD on our test set: it tends to find isolated structures in relatively few graphs as anomalous.  

We test this method using the 80-20 split procedure used to test our approach outlined in Section 5.1: a random sampling of 80\% of the login graphs are tested at a time, and we repeat 25 times (the rationale is the same as before---to test generalizability against different login histories).  The computed $\overline{{\rm FPR}}$ varies considerably across users, from a maximum of 60\% down to 0\%, with a mean of 6.5\%.  

To measure true positive rates, we perform the above procedure for each adversarial graph type, $i$, injecting an adversarial graph into a different randomly selected login graph in each of the 25 iterations.  We find this technique fails to detect {\it any} novel-to-novel adversarial prototypes. This is likely because novel-to-novel logins appear as separate components in their parent graph, and subgraphs of similar structure might not appear elsewhere in $\mathcal{G}^u$.  While this might sound precisely like the kind of novel substructure to trigger an anomaly, GBAD looks for anomalous substructure {\it modifications}, and so requires that a similar subgraph exist as a normal pattern across $\mathcal{G}^u$. 

Interestingly, GBAD performs better against the more difficult novel-to-known login types.  The likely reason is that malicious logins now include known systems, and so modify existing substructures, either by insertion or more generally.  These are the kinds of anomalies that GBAD is better suited to detect.  Unfortunately, it still performs rather poorly, with an average $\overline{{\rm TPR}}_i$ of only 11\%. 

\subsection{Vertex anomalies with one-class learning}
We next compare the reconstruction error-based anomaly detection method of our approach with one based on one class learning.  For the one class learning-version of our approach, we train a classifier directly on the graph features of Table 1: there is no compression step.  We select a one-class support vector machine (OCSVM) with a Gaussian kernel.   We perform multiple rounds of 5-fold cross validation on each user's complete set of login graphs: if a tested vertex is novel and found to be anomalous, this constitutes a false positive.  To test adversarial graph detection, we perform the same cross validation separately for each adversarial graph type, injecting a graph into a randomly selected test login graph in each fold.   We plot the receiver operating characteristic (ROC) curves for novel-to-novel detections for our reconstruction-error based method using NMF and PCA, and the OCSVM based method.  

The $\overline{{\rm FPR}}$ of each user model is averaged to obtain the FPR values, and the $\overline{{\rm TPR}}_i$ for each adversarial graph $i$ for each user are averaged to obtain the TPR values.  The PCA-based method performs best, as expected from our previous results, and the OCSVM performs worst.  This is an interesting result: in using the full 13-dimensional graph feature space, the OCSVM model is most expressive; however, it does not discriminate as well as the lower-dimensional NMF- and PCA-based ensembles.  Evidently, feature compression followed by suitable ensembling results in a more expressive model in this case.      
\begin{figure}[ht]
\centering
\includegraphics[width=3.5in]{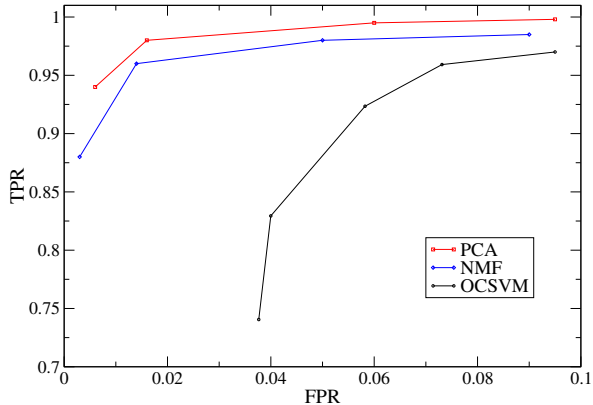}
\caption{\footnotesize{ROC curves of the NMF and PCA reconstruction error-based models and the OCSVM.}}
\label{ROC}
\end{figure}
\subsection{Global anomalies}
Yet another approach to this problem is to assess individual login graphs within a user's history as anomalous or normal, rather than individual vertices or substructures within them as explored previously.  The graph classification method proposed in \cite{Li12} evaluates graphs across a large set of global measures and trains a support vector machine to classify them.  We can apply this approach to test if the login graph $G^u_0$ is anomalous by learning the user's historical sequence, $\mathcal{G}^u - \{G^u_0\}$.  Following \cite{Li12}, we evaluate 17 measures for each login graph: average degree, average clustering coefficient, average/maximum/minimum effective eccentricity, average closeness centrality, percentage of central points (vertices with eccentricity equal to the graph diameter), giant connected ratio, percentage of isolated points, percentage of end points (vertices with $k = 1$), number of nodes, number of edges, spectral radius (magnitude of the largest eigenvalue of the adjacency matrix, $|\lambda_1|$), second-largest eigenvalue, trace (the sum of all eigenvalues, $\sum_i \lambda_i$), number of eigenvalues, and the energy ($E_\lambda = \sum_i \lambda_i^2$).  The remaining three features from \cite{Li12} require vertex attributes, which our login graphs lack.  To determine $\overline{{\rm FPR}}$, we perform leave-one-out cross validation on $\mathcal{G}^u$ so that each graph $G^u_i$ has an opportunity to be the test graph, $G^u_0$.  This is a good way to test the generalizability of the method. To determine $\overline{{\rm TPR}}_i$, we perform leave-one-out cross validation on $\mathcal{G}^u$ and inject adversarial graph $i$ into each $G^u_0$, for each graph $i$.  

\begin{figure}[ht]
\centering
\includegraphics[width=5in]{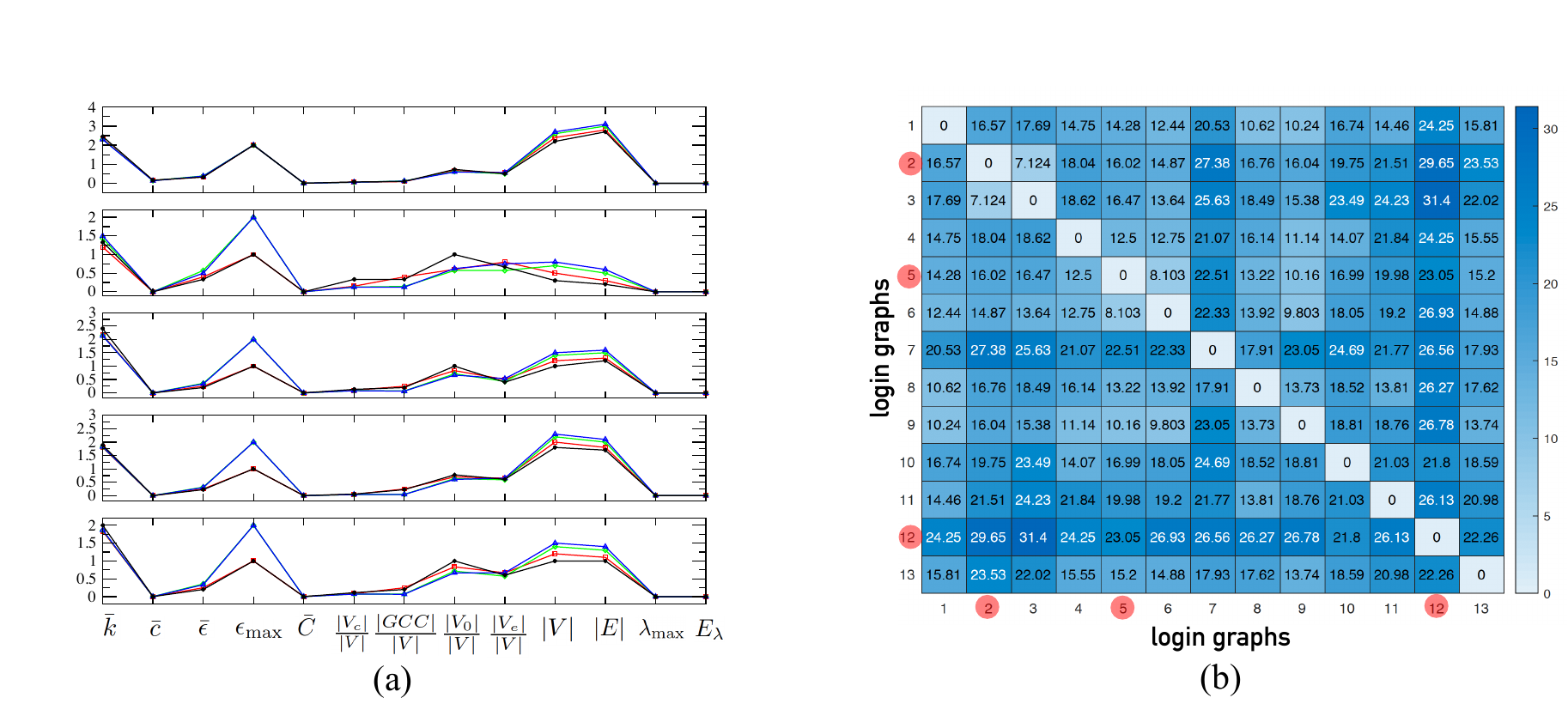}
\caption{\footnotesize{(a) Five Domain Administrator login graphs evaluated across the measures of \cite{Li12}: normal (black), and with type 1 (red), type 5 (green), and type 14 (blue) adversarial graphs embedded. The values of $|V|$ and $|E|$ were each reduced by a factor of 10 in these plots for reasons of scalability. (b) Similarity matrix of 13 Domain Administrator login graphs computed using pair-wise Canberra distance.  A type 14 adversarial graph was embedded in login graphs 2, 5, and 11 to illustrate its effect.}}
\label{global}
\end{figure}
Using an SVM with linear kernel, we tune the classifier to achieve true positive rates comparable to our method; we find that the best average $\overline{{\rm FPR}}$ over all users is high, around 25\%.  The average $\mu(\overline{{\rm TPR}}_i)$ for novel-to-novel logins is 94\% and against novel-to-known is 92\%.  While the measures proposed in \cite{Li12} were devised to be a fairly complete and general set of global graph properties, the adversarial graphs injected into normal login graphs tend to constitute only a minor perturbation to these properties.  In Figure \ref{global} (a), the global measures of the first five login graphs of the example Domain Administrator are plotted (top to bottom).  The four data series in each plot show the values for the original login graph (black), and the login graph with a type 1 (red), type 5 (green), and type 14 (blue) adversarial graph embedded.  Aside from $\epsilon_{\rm max}$, the measures scarcely budge in most graphs. This makes them hard to detect reliably without incurring increased false positives. 

As a final comparison, we consider global anomaly detection based on the metrics of Table 1, which we have seen are discriminating for the special case of adversarial lateral movement.  Following \cite{Berlingerio}, we summarize each login graph according to the mean, variance, skewness, and kurtosis of each graph measure taken over all nodes in the graph.  In \cite{Berlingerio}, a distance-based anomaly detection scheme using the Canberra distance, $d({\mathbf x},{\mathbf y}) = \sum_i |{\mathbf x}_i - {\mathbf y}_i|/({\mathbf x}_i+{\mathbf y}_i) $, is implemented. The authors select this metric due to its discriminative power for small differences near zero.  Each login graph is described by a $4\times 13 = 52$-dimensional feature vector (for the 4 statistical moments of the 13 graph measures), and distances are computed pairwise among all vectors.  Anomaly detection can then be done by imposing a threshold on the average distance of a sample from all others.  The sensitivity and false positive rate of the detector can be tuned by varying this threshold; however, we find that performance is considerably worse than the SVM-based approach of \cite{Li12} regardless of threshold.  Mean $\overline{{\rm FPR}}$ varies between 15\% and 27\%, with mean $\mu(\overline{{\rm TPR}}_i)$ for novel-to-novel logins lying between 53\% and 65\%, and for novel-to-known logins between 47\% and 54\%.  The similarity matrix formed from the 13 login graphs of the example Domain Administrator is shown in Figure \ref{global} (b), where adversarial graph type 14 was embedded into login graphs 2, 5, and 12 to reveal its effect on the distance measure.  The effect is indeed small, for this user at least, offering an illustration of why this method fails to reliably discriminate malicious from normal novel logins.

Table \ref{tab2} summarizes the performances of the methods explored in this section.  While our approach out-performs the others, this is not because it is generally better, but rather that it was designed specifically with our application in mind.  Discovering stealthy malicious logins on daily timescales is a difficult problem that general-purpose graph-based anomaly detection methods struggle with.  These results confirm the need for a novel methodology to address this problem.  

\begin{table}
\begin{center}
\begin{tabular}{|c|c|c|c|}
\hline
Method& mean $\overline{{\rm FPR}}$&\multicolumn{2}{c|}{mean $\mu(\overline{{\rm TPR}}_i)$}\\
\hline
Local anomalies, NMF recon. error &2.5\%&96\%&91\% \\
Local anomalies, PCA recon. error &2.2\%&98\%&96\% \\
Local anomalies, OCSVM&6\%&93\%&88\% \\
Substructure anomalies, GBAD \cite{Eberle07} &6.5\%&0\%&11\% \\
Global anomalies, aggregated graphs \cite{Kent} &59\%&-&- \\
Global anomalies, OCSVM \cite{Li12} &25\%&94\%&92\% \\
Global anomalies, distance-based \cite{Berlingerio} &23\%&62\%&53\% \\
\hline
\end{tabular}
\end{center}
\caption{\footnotesize{Performance of comparable methods explored in this section. The first column under mean $\mu(\overline{{\rm TPR}}_i)$ gives performance for novel-to-novel logins, and the second for novel-to-known logins.}}
\label{tab2}
\end{table}

\section{Conclusions}
We have presented a new capability that determines whether a user's novel logins are potentially malicious.  We have cast this approach as a graph anomaly detection problem, where individual vertices across a collection of graphs representing a user's login history are tested for anomalous behavior.  The method makes use of local graph measures to characterize each vertex and uses the reconstruction error of a compression transformation applied to these features to identify outliers.  

The method is tested on a set of 78 high-privileged users with login data from a real, operational enterprise network.  Models are developed using four weeks of historical login data for each user for two use-cases: one focused on detecting malicious logins among only novel systems, and one focused on detecting malicious logins among novel systems and a single known system, both over a 24-hour period.  The average false positive rates across all users for these two cases were 1.6\% and 2.2\%, and the average true positive rates, as assessed by validating against a set of simulated adversarial login graphs, were 98\% and 96\%, respectively.  False positive rates are taken with respect to all {\it novel} systems accessed by the user on the given day, not {\it all} systems.  For most users in our test set, novel logins were relatively rare, making up typically less than 5\% of a given user's daily logins.  Therefore, in comparison to basic detectors based solely on novelty, this approach alerts on significantly fewer logins.  

This capability is intended to be deployed operationally for use-cases much like that demonstrated here: models constructed for $\mathcal{O}(10)$ high-value accounts and with daily login monitoring using authentication logs.  Even with relatively low false positive rates, for many dozens of accounts with daily testing, the number of false alarms might be significant.  We therefore envision this kind of detection capability constituting one part of a larger correlation program (like, {\it e.g.} \cite{milajerdi}), in which suspicious login events are considered together with other indicators to help shape a threat picture. Correlation-based detection paradigms are more tolerant of false positives, and less-discerning sensors provide the correlator with more potentially useful input.  Our method can be tuned to adjust false positives by considering different priorities during model construction, and by tuning the detection threshold, $\alpha$.  

Though we have presented a model based on a specific set of 13 graph measures, this approach is not wedded to any particular feature set.  Indeed, measures can be selected according to the use-case at hand: for authenticated lateral movement, we find the current set to be useful but other measures might be more suited to discovering other malicious network activities, like scanning, denial of service, or malware propagation. Further, this approach is not wedded to the use of NMF or PCA; in fact, any compressive function, like more general autoencoders, might be used and adapted to the application at hand. 

The present capability has been developed with malicious novel logins in mind, but it might be of interest to expand this method to discover anomalous logins to known systems.  This kind of analysis might closely resemble the previously discussed approaches to anomaly detection in dynamic networks, in which the properties of vertices are tracked over time for unusual changes.  This has implications for the compression function that is used: non-convex local optimization-based methods like NMF are not guaranteed to yield the same role assignments at each time step, making them unsuitable for application to time series.    

Further research might also be done on more complex models than those investigated here.  In this work we only exhaustively searched 2- and 3-dimensional models, since an extensive analysis of optimal models quickly becomes prohibitive ({\it e.g.} a grid search over 4-parameter models would involve $\binom{13}{4} = 715$ evaluations for each user, a factor of 10 larger than the 2-parameter model space explored here.)  Indeed, better-performing models than those found here might lie in these higher-dimensional spaces.   Nevertheless, the models constructed in this study are shown to be greatly more effective than general-purpose graph anomaly detection schemes at identifying stealthy, malicious logins over short, daily timescales.  

\section{Acknowledgement}
The author acknowledges use of the GBAD software from \\\url{https://users.csc.tntech.edu/~weberle/gbad/}.   
\section{References}
\bibliography{powell}

\end{document}